\documentclass{article}
\PassOptionsToPackage{numbers}{natbib}

\usepackage[preprint,main]{NeurIPS}
\usepackage[utf8]{inputenc} 
\usepackage[T1]{fontenc}    
\usepackage{hyperref}       
\usepackage{url}            
\usepackage{booktabs}       
\usepackage{amsfonts}       
\usepackage{nicefrac}       
\usepackage{microtype}      
\usepackage{xcolor}         
\usepackage{amsmath}
\usepackage{amssymb}
\usepackage{graphicx}
\usepackage{algorithm}
\usepackage{algorithmicx}
\usepackage[noend]{algpseudocode}
\usepackage{subcaption}
\usepackage{mathtools}
\usepackage[normalem]{ulem}

\newcommand{\nop}[1]{}

\title{Adaptive Scheduling Pipeline for Multi-Instance Asynchronous Reinforcement Learning}

\author{%
  Salah Chikhi\thanks{Equal contribution.} \\
  École polytechnique
  \And
  Luis Zuin\footnotemark[1] \\
  Huawei Technologies
  \And
  Entong Li \quad Li Zeng\thanks{Corresponding author: \texttt{zengli@sribd.cn}.} \\
  Shenzhen Research Institute of Big Data
}

\begin{document}

\maketitle

\begin{abstract}

We present the Adaptive Scheduling Pipeline (ASP), an offline rollout scheduler for multi-instance asynchronous reinforcement learning (RL) of large language models (LLMs).
ASP jointly reduces three sources of rollout underutilization---padding bubbles from intra-minibatch length heterogeneity, skewness bubbles from long-tail outputs, 
and sorting bubbles from fluctuations in training-queue occupancy induced by output-length-sorted dispatch---through a single offline preprocessing pass. 
This enables efficient multi-instance rollout that overcomes the early scaling plateau of single-instance inference 
while reducing fluctuations in the supply of generated samples to the trainer.
We integrate the scheduler into an asynchronous architecture that decouples rollout from training while keeping policy lag observable and bounded.
Within our asynchronous architecture, ASP's scheduling reduces mean iteration time by 10.9\%. The complete system improves end-to-end throughput by up to $3.22\times$ and reduces iteration time by up to 76.8\% relative to MindSpeed-RL, while reducing iteration time by up to 62.5\% relative to verl. It also scales more robustly with additional generation resources than these baselines and shows no visible convergence degradation in the measured bounded-lag regime.
ASP has since been used in an internal telecom-domain RL post-training workflow, reducing the reported training period from more than three months to two months and producing a domain-enhanced model for knowledge answering and tool calling.
\end{abstract}

\section{Introduction}
\label{sec:introduction}

Reinforcement learning (RL) has become a central component of large language
model (LLM) post-training, improving alignment and reasoning by training on
generated responses and reward signals~\cite{ouyang2022training,havrilla2024teachinglargelanguagemodels}.
However, RL post-training also creates a demanding systems problem: each policy
update requires rollout generation, scoring, log-probability computation, and
backpropagation. As models produce longer outputs, rollout/generation often
becomes the dominant bottleneck, accounting for a large fraction of
training time in modern RL with Verifiable Rewards (RLVR) systems~\cite{wen2025rlvr, openrlhf}.

Large-scale RL for LLMs is commonly executed in a \emph{hybrid} loop: for each
batch, the system first performs inference and then runs a training step on the
resulting samples. This sequential organization is simple and keeps rollouts
close to the current policy, but it has several drawbacks. First, it creates
idle time: while the trainer updates, no new samples are generated. Second, it
forces repeated mode switches and possible re-sharding on the same machines.
Third, it couples two stages with different system profiles: backpropagation is
typically compute-bound and benefits from synchronized parallelism, whereas
autoregressive generation is memory-bandwidth-bound and often reaches an early
scaling plateau when more resources are added to a single inference
instance~\cite{streamRL}.

Asynchronous and disaggregated pipelines address part of this problem by allowing rollout and training to proceed concurrently~\cite{streamRL,asyncflow,noukhovitch2025asynchronousrlhffasterefficient}. In this producer--consumer setting, generation instances produce samples that the trainer consumes for policy updates, with a queue buffering completed samples until they are needed. However, separation alone does not overcome the early scaling plateau of single-instance rollout. Scaling rollout across multiple generation instances exposes a local--global scheduling conflict: decoding minibatches should use each instance efficiently despite variable response lengths, while their dispatch order should maintain a steady supply of samples to the trainer over time.

We identify three sources of underutilization in this pipeline. The first is \emph{padding bubbles}: prompts in the same inference minibatch produce responses of different lengths, so shorter sequences finish early while the generation actor remains occupied by the longest generations. The second is \emph{skewness bubbles}, following StreamRL~\cite{streamRL}: LLM output lengths are often heavy-tailed, and a few long generations can collapse the effective active batch size near the end of decoding. Sorting prompts by predicted output length helps reduce both local inefficiencies, but can create a global imbalance between generation and training. If short minibatches are dispatched first and progressively longer minibatches later, the sample arrival rate decreases over time: the queue may initially build up and then drain as longer minibatches dominate. Queue buildup makes samples wait through additional policy updates, increasing policy lag; if the queue later empties, the trainer must wait for new samples. We call this temporal imbalance caused by sorted dispatch \emph{sorting bubbles}.

We propose the \textbf{Adaptive Scheduling Pipeline (ASP)}, an offline rollout scheduler for multi-instance asynchronous RL. ASP first uses predicted output lengths and dynamic programming to construct length-homogeneous decoding minibatches, improving local rollout efficiency. It then reorders these minibatches into consecutive batch groups with balanced aggregate predicted load, smoothing the supply of samples to the trainer over time. This single offline preprocessing pass uses predicted lengths to jointly target padding, skewness, and sorting bubbles without adding online scheduling overhead during rollout.

ASP is integrated into a multi-instance asynchronous RL pipeline designed to keep policy lag bounded.
A Training Unit
(TU) performs reward computation, log-probability evaluation, and policy
updates; one or more Generation Units (GUs) host multiple generation actors that perform rollouts in parallel; and
a Control Unit (CU) dispatches scheduled minibatches and forwards
completed samples through a controlled Training Queue. The queue absorbs
short-term variability and provides backpressure, while periodic weight refresh
keeps the rollout policy close to the trainer. This design preserves the main
benefits of disaggregation---continuous rollout, multi-instance scaling, and
independent training/generation configurations---while keeping policy lag
observable and bounded.


ASP targets a complementary scheduling layer relative to prior asynchronous
systems. StreamRL~\cite{streamRL} uses an output-length ranker primarily for
inter-instance routing, assigning the top $\alpha$\% longest predicted samples
to dedicated instances and prioritizing long requests to reduce generation
makespan. It does not optimize which prompts share a decoding minibatch or how
the resulting minibatches are ordered to stabilize trainer arrivals.
AsyncFlow~\cite{asyncflow} instead focuses on asynchronous dataflow across RL
stages, scheduling ready samples across tasks and data-parallel consumers with
consumer-side token balancing; its mechanisms do not construct
predicted-length decoding minibatches or balance their dispatch order.
Other recent systems address complementary aspects of asynchrony: AReaL~\cite{fu2026areallargescaleasynchronousreinforcement}
controls staleness through rollout admission and interrupted generation,
whereas Laminar~\cite{sheng2025laminarscalableasynchronousrl} reactively handles in-flight long-tail trajectories.
ASP acts before rollout, jointly choosing predicted-length minibatch composition
and balancing consecutive batch groups so that local reductions in padding
and skewness do not create unstable trainer arrivals. Thus, ASP is
complementary to disaggregation, routing, and staleness-control mechanisms
rather than an alternative to them. 


We implement ASP inside MindSpeed-RL~\cite{MindSpeed-RL} with Group Relative Policy Optimization (GRPO)~\cite{deepseekmath} on Ascend 910B3 Neural Processing Units (NPUs), and evaluate it on the Qwen3 Mixture-of-Experts (MoE) and dense models~\cite{qwen3-techreport}.
\nop{
and evaluate it on Qwen3 MoE and dense models~\cite{qwen3-techreport}.
ASP improves end-to-end throughput by up to $3.22\times$ (+222\%), reduces iteration time by up to 76\%, improves scheduling ablation metrics, and shows no visible convergence degradation under a measured off-policy distance of approximately 3.3 trainer steps. 
}
Our contributions are:
\vspace{-0.1in}
\begin{enumerate}
    \item We identify a local--global scheduling conflict in multi-instance rollout and propose ASP, an offline scheduler that combines predicted-length minibatch construction with balanced batch-group ordering to reduce padding, skewness, and sorting bubbles.
    \item We isolate ASP's scheduling contribution through staged scheduling ablations and predictor-quality analysis, showing that naive length sorting improves local efficiency but destabilizes trainer arrivals, while balanced ordering preserves the local gain and restores queue stability.
    \item We integrate ASP into a bounded-lag asynchronous RL pipeline and evaluate it on Qwen3 MoE and dense models~\cite{qwen3-techreport} against matched hybrid and optimized asynchronous/disaggregated baselines.
\end{enumerate}
\vspace{-0.1in}
\section{Background and related work}
\label{sec:background}

\vspace{-0.05in}
\subsection{RL post-training pipelines}

Large-scale RL post-training for LLMs is commonly implemented as a hybrid loop: a batch of prompts is first sent to rollout/generation, the generated samples are scored, and the resulting data are then consumed by the trainer for log-probability evaluation and policy update. This sequential organization is simple, but it couples two stages with different system profiles. This coupling can leave accelerators idle, force repeated mode switches between inference and training, and prevent each stage from using its most suitable parallelism strategy.

Instead, asynchronous pipelines overlap rollout with training, while disaggregation allows the two stages to use separate resources and configurations~\cite{streamRL,asyncflow,noukhovitch2025asynchronousrlhffasterefficient}. This requires coordinating the delivery of generated samples to the trainer and the refresh of model weights used for generation. With multiple generation instances, efficient execution also depends on allocating resources between stages and scheduling variable-length rollouts to limit idle time~\cite{streamRL,asyncflow}.

\vspace{-0.05in}
\subsection{Existing asynchronous systems and remaining gaps}

Prior asynchronous RL systems address complementary layers of rollout
execution. StreamRL~\cite{streamRL} improves disaggregated RL training through
stream generation to reduce pipeline bubbles and an output-length ranker
to mitigate skewness bubbles. It routes the top $\alpha$\% of predicted
long-tail prompts to dedicated machines, improving batching efficiency.
While closely related to our motivation, StreamRL primarily uses length
prediction for inter-instance routing, without explicitly optimizing decoding-minibatch composition. 
Moreover, $\alpha$ is fixed, and its dispatching policy does not explicitly balance the predicted load of consecutive batch groups to smooth sample arrivals at the trainer.
In contrast, ASP uses predicted lengths offline to construct 
length-homogeneous minibatches and reorder them into balanced batch
groups, jointly addressing padding, skewness, and sorting bubbles.

AsyncFlow~\citep{asyncflow} uses TransferQueue for asynchronous dataflow
across RL stages and tasks, with consumer-side token balancing. Its padding optimizations target data storage and transmission rather than decoding minibatch composition or dispatch ordering. Other systems address different bottlenecks:
AReaL~\citep{fu2026areallargescaleasynchronousreinforcement} controls
staleness through generation admission and interrupted decoding across
weight updates; Laminar~\citep{sheng2025laminarscalableasynchronousrl}
reactively relocates in-flight long-tail trajectories. ASP instead builds
its schedule before rollout, without interrupting or migrating work that
has already been dispatched. These mechanisms are complementary rather
than mutually exclusive.



verl~\cite{verl,hybridflow} is an RL framework for LLM
post-training supporting disaggregated rollout and training. It notably supports training-side remove-padding and dynamic batching,
as well as rollout-side continuous batching. However, unlike ASP, it does not
explicitly optimize decoding-minibatch composition or batch-group
ordering to stabilize trainer arrivals. This makes it a natural asynchronous baseline. A more detailed comparison with commonly used RL frameworks is provided
in Appendix~\ref{app:extended_related_work}. Table~\ref{tab:related-comparison} summarizes this design space. 

\vspace{-0.05in}
\begin{table}[htbp]
  \centering
  \caption{Comparison of ASP with existing asynchronous RL systems.}
  \label{tab:related-comparison}
  \footnotesize
  \setlength{\tabcolsep}{3pt}
  \begin{tabular}{lcccc}
    \toprule
    \textbf{Method} &
    \textbf{Async} &
    \shortstack{\textbf{Multi-inst.}\\\textbf{rollout}} &
    \shortstack{\textbf{Padding-aware}\\\textbf{rollout scheduling}} &
    \shortstack{\textbf{Sorting-bubble}\\\textbf{mitigation}} \\
    \midrule
    Hybrid RL & No & No & No & No \\
    StreamRL~\citep{streamRL} & Yes & Yes & Partial & No \\
    AsyncFlow~\citep{asyncflow} & Yes & Config.-dep. & No & No \\
    AReaL~\citep{fu2026areallargescaleasynchronousreinforcement} & Yes & Yes & No & No \\
    Laminar~\citep{sheng2025laminarscalableasynchronousrl} & Yes & Yes & No & No \\
    verl~\citep{verl} & Yes & Config.-dep. & No & No \\
    ASP & Yes & Yes & Yes & Yes \\
    \bottomrule
  \end{tabular}
\end{table}

\vspace{-0.1in}

\section{Offline scheduling for multi-instance rollout}
\label{sec:scheduling}

In ASP, scheduling is not only a throughput optimization for rollout. It is also the mechanism that makes multi-instance asynchronous generation more predictable:
the same strategy that improves per-minibatch utilization also smooths the
producer--consumer dynamics seen by the Training Queue over time. 
This section describes the scheduling problem and the three-step offline preprocessing pipeline used by ASP. Table~\ref{tab:asp-scheduling-pipeline} summarizes the role of each step before we describe the steps in detail. 

\vspace{-0.1in}
\begin{table}[htbp]
  \centering
  \caption{ASP scheduling pipeline. Each offline step targets a different source
  of rollout inefficiency or queue instability.}
  \label{tab:asp-scheduling-pipeline}
  \small
  \begin{tabular}{lll}
    \toprule
    \textbf{Step} & \textbf{Mechanism} & \textbf{Main effect} \\
    \midrule
    1 & Output-length prediction & Orders prompts by expected rollout length \\
    2 & Dynamic programming minibatch construction & Reduces intra-minibatch length heterogeneity \\
    3 & Balanced batch-group ordering & Smooths queue arrivals over time \\
    \bottomrule
  \end{tabular}
\end{table}
\vspace{-0.1in}

We use the term \emph{sample} for one prompt--response trajectory, \emph{minibatch} for the set of prompts decoded in parallel by one generation actor, and \emph{global batch} for the set of prompts consumed in one training iteration.
Each prompt is sampled several times, and all its samples are generated in the same minibatch and consumed in the same global batch; batch sizes are therefore counted in prompts.
During rollout, each generation actor repeatedly receives a minibatch, generates the corresponding responses, and returns the completed samples to the Training Queue only after the entire minibatch finishes. The
scheduler decides how prompts are grouped into minibatches and how those
minibatches are ordered before dispatch. Its objective is twofold: reduce local underutilization during generation, and make the arrival rate of generated samples at the Training Queue balanced over time. 

\vspace{-0.05in}
\subsection{Bubble sources}

Rollout efficiency is limited by three sources of underutilization. The first is
\emph{padding bubbles}: prompts in the same minibatch generate responses of
different lengths, so shorter sequences finish earlier while the actor remains
occupied by the longest ones. 
As a result, useful token generation can occupy only a small fraction of the token slots implied by the minibatch size and longest response.

The second is \emph{skewness bubbles}, following StreamRL~\cite{streamRL}: LLM
response lengths are often heavy-tailed, and a few very long generations can
collapse the effective active batch size near the end of decoding. Sorting
prompts by predicted output length helps reduce both padding and skewness within a minibatch, because prompts with similar expected lengths are decoded together.

The third source is specific to asynchronous rollout. We call it
\emph{sorting bubbles}. If the dataset is simply sorted by predicted length,
early minibatches tend to be short and later minibatches tend to be long. This
improves local length homogeneity, but can destabilize the global pipeline:
generation may initially overproduce samples and fill the Training Queue, then
slow down as longer minibatches dominate, causing the queue to drain. Thus,
naive length sorting can trade local efficiency for temporal instability. ASP
addresses this by separating two goals: first forming length-homogeneous
minibatches, then reordering those minibatches into balanced batch groups.
\vspace{-0.05in}
\subsection{Step 1: output-length prediction}

ASP first predicts the output length of each prompt and sorts the dataset by the predicted values. Note that exact output length predictions are neither required nor attainable: response lengths vary across generations for the same prompt and change as the model is updated during RL training. The predictor only needs to be roughly accurate, producing values close enough to the actual lengths and an ordering that separates prompts whose expected response lengths differ substantially.


In our implementation, we collect prompt--length pairs by running inference with the target model on a small subset of prompts, then fine-tune Qwen~2.5 1.5B~\cite{qwen25-techreport} to predict output lengths (Appendix~\ref{app:output_length_predictor}). We measure its ranking quality with Kendall's $\tau$~\cite{kendall1938rank}, and Section~\ref{sec:scheduling_ablation} evaluates scheduling across prediction qualities. The predictor remains fixed throughout training, a choice supported by the limited response-length drift in the main \texttt{Qwen3-30B-A3B} run, where mean length changes by only 1.6\% (Appendix~\ref{app:scheduling_runtime_traces}).


Our use of predictions differs from StreamRL's output-length ranker. 
StreamRL uses predicted lengths to route long-tail samples to dedicated instances and prioritize longer requests. As we explain in the following steps, ASP uses predicted lengths offline, once before training, to construct minibatches and balance consecutive batch groups.

\subsection{Step 2: length-homogeneous minibatch construction}

After sorting prompts by predicted length, ASP constructs inference minibatches. With a fixed minibatch size, large gaps in predicted output lengths may fall within a minibatch, forcing prompts of substantially different output lengths to be decoded together. To mitigate this, we allow minibatch sizes to vary slightly between a base size $b$ and $b+\kappa$, where $\kappa$ is typically small.

Let $l_0 \leq \cdots \leq l_{n-1}$ be the sorted predicted output lengths. A
partition of the dataset is defined by endpoints
$0=x_0 < x_1 < \cdots < x_{N_B}=n$, where minibatch $i$ contains prompts
$x_i,\ldots,x_{i+1}-1$. For each minibatch, we define the utilization proxy:
\begin{equation}
R_i =
\frac{\sum_{j=x_i}^{x_{i+1}-1} l_j}
     {(x_{i+1}-x_i)\, l_{x_{i+1}-1}}
    \label{eq:utilization_proxy}
\end{equation}

This ratio compares the predicted useful generated tokens with the total token slots implied by the minibatch size and longest predicted output.
Hence, we define $1-R_i$
to be the estimated padding bubble rate of minibatch $i$, and
\begin{equation}
\bar{P} = 1 - \frac{1}{N_B}\sum_{i=0}^{N_B-1} R_i
    \label{eq:padding_bubble_rate}
\end{equation}

is the average predicted padding bubble rate, which should be minimized. With $\kappa$ small relative to $b$, $N_B$ varies little over the feasible set, motivating the use of summed utilization as a surrogate objective. In practice, ASP therefore maximizes $\sum_{i=0}^{N_B-1}R_i$ subject to $b \leq x_{i+1}-x_i \leq b+\kappa$. This optimization is solved by dynamic programming in $O(\kappa n)$ time and $O(n)$ space. The full recurrence and pseudocode are given in Appendix~\ref{app:scheduling_details}.

Hence, ASP preserves the benefits of output length sorting while reducing within-minibatch output length disparities through flexible minibatch sizing. The
result is a sequence of minibatches that are locally efficient for generation,
but still ordered from short to long. The next step tackles this
temporal imbalance.

\subsection{Step 3: balanced batch-group ordering}

The final step reorders minibatches to balance the predicted load of consecutive batch groups and stabilize arrivals at the Training Queue.
Let $B_i$ be the predicted load of minibatch $i$, defined
as the sum of predicted output lengths in that minibatch. ASP groups
minibatches into batch groups of size $k$, and seeks a permutation $\sigma$ such
that each group has approximately the same total predicted load. With $G = \lceil N_B/k \rceil$ groups and target group
load $M = \frac{1}{G}\sum_{i=0}^{N_B-1} B_i$,
defined as the average predicted load per group,
the objective is
\begin{equation}
    \min_{\sigma}
\frac{1}{G}\sum_{g=1}^{G}
\left(
\sum_{i \in \mathcal{G}_g} B_{\sigma(i)} - M
\right)^2
\label{eq:step3_objective}
\end{equation}

where $\mathcal{G}_g$ denotes the indices assigned to group $g$. This objective
directly targets sorting bubbles: instead of dispatching all short minibatches
first and all long minibatches later, ASP mixes short and long minibatches so
that consecutive groups have similar predicted processing time.

We solve this reordering problem with simulated annealing (SA)~\cite{kirkpatrick1983optimization}
in our implementation. After reordering, batch groups may
still be shuffled for training randomness, while preserving their internal load
balance. The detailed objective, simulated annealing procedure, and
hyperparameter choices are deferred to Appendix~\ref{app:scheduling_details}.

Overall, ASP performs one offline preprocessing pass: predict lengths, build
length-homogeneous minibatches, and reorder them into balanced batch groups.
This produces rollout work units that are efficient for individual generation
actors and stable for the global Training Queue, which is precisely the
scheduling layer missing from naive asynchronous disaggregation.

\section{Integration into bounded-lag asynchronous RL}
\label{sec:integration}

The scheduling pipeline described above is designed to run inside an
asynchronous RL system where rollout and training proceed concurrently. 
This section summarizes the system integration needed to make ASP practical: the architecture that consumes the scheduled minibatches, and the bounded-lag
weight-refresh mechanism that keeps asynchronous training close to the current
policy.

    

\vspace{-0.1in}
\begin{figure}[htbp]
    \centering
    \includegraphics[width=0.8\linewidth]{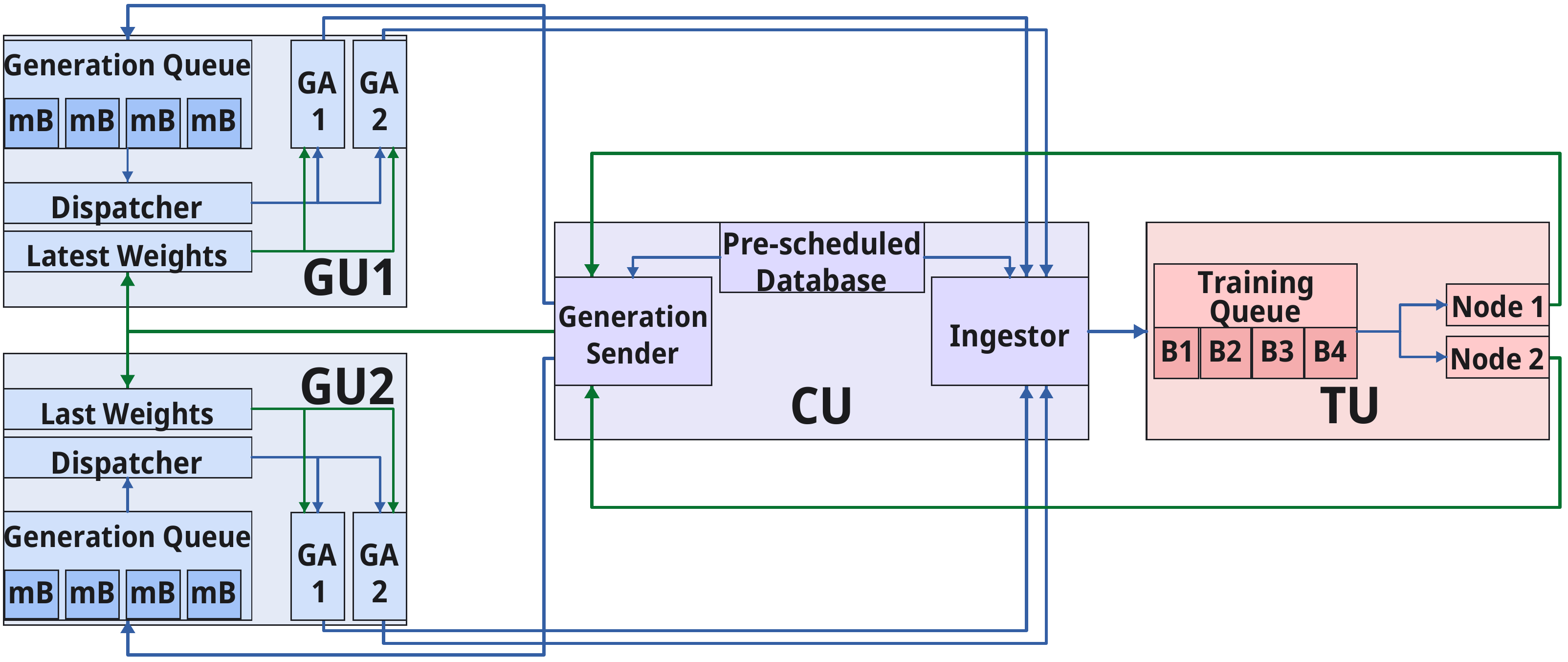}
    \caption{ASP architecture. Control Unit dispatches scheduled minibatches to Generation Units, receives completed samples, and forwards them to Training Queue consumed by Training Unit. GA denotes a generation actor, mB a minibatch, and B a global batch.}
    
    \label{fig:app_asp_architecture}
\end{figure}
\vspace{-0.1in}

\subsection{Architecture overview}

ASP is integrated into a decoupled RL post-training pipeline with three components: a \emph{Training Unit} (TU), one or more \emph{Generation Units} (GUs), and a \emph{Control Unit} (CU) co-located with the TU.
The TU performs reward
computation, log-probability evaluation, and policy updates. Each GU hosts
multiple generation actors, typically vLLM~\cite{kwon2023efficientmemorymanagementlarge}
instances, which repeatedly receive scheduled inference minibatches, generate
responses, and return completed samples. The CU coordinates this
producer--consumer interaction by dispatching minibatches to the GUs and forwarding completed samples to the TU through a Training Queue. 
Figure \ref{fig:app_asp_architecture} gives the system diagram and more details are given in Appendix~\ref{app:system_details}.

This architecture serves two purposes. First, it allows rollout and training to
operate continuously instead of alternating as in hybrid pipelines. Second, it
keeps the two stages independently configurable: the TU and GUs may use
different parallelism layouts, batch sizes, and hardware allocations. This
independence is important because backpropagation and autoregressive generation
have different system profiles, and a single shared configuration is often
suboptimal for at least one of them.

The Training Queue is the main stability mechanism. It buffers short-term fluctuations in generation time and allows the CU to monitor queue occupancy for backpressure control. 
When generated samples arrive faster than the TU consumes them, the CU can pause the start of new decoding minibatches; actors already decoding finish their current minibatches. Generation resumes as the Training Queue drains.
ASP's scheduler reduces the variance of generated-sample arrival times
before execution, while the queue provides online stabilization during training.
Detailed CU/TU/GU implementation choices 
and the resource allocation heuristic are deferred to
Appendix~\ref{app:system_details}.

\subsection{Weight refresh and bounded policy lag}

Beyond balancing rollout and training rates, asynchronous execution introduces policy lag: samples may be generated by older model weights than those currently being trained, potentially destabilizing learning if the lag becomes too large.
ASP therefore uses a bounded-lag weight-refresh path. The TU exports weight snapshots at a configurable interval, set to one training step in all our experiments. Because the TU and GUs may use different parallelism layouts, this snapshot is asynchronously converted into an inference-ready format before being loaded by generation actors. Finally, actors reload weights asynchronously: if an actor is already decoding a minibatch, it finishes that minibatch before reloading, avoiding a global pause of all generation workers.

This design preserves the main benefit of disaggregation---continuous
generation---while keeping the lag observable and bounded. For a minibatch that
arrives at the Training Queue, the effective off-policy distance, measured in training steps, can be
approximated by
\vspace{-0.02in}
\begin{equation}
d_{\mathrm{off}}
\approx
Q_{\mathrm{arrival}}
+
\frac{
T_{\mathrm{save}} + T_{\mathrm{convert}} + T_{\mathrm{load}} + T_{\mathrm{gen}}
}{
T_{\mathrm{train}}
}
    \label{eq:d_off}
\end{equation}

\vspace{-0.03in}
where $Q_{\mathrm{arrival}}$ is the Training Queue depth measured in global batches immediately before the minibatch is inserted, $T_{\mathrm{save}}$, $T_{\mathrm{convert}}$,
and $T_{\mathrm{load}}$ are the snapshot export, conversion, and reload times,
$T_{\mathrm{gen}}$ is the generation time of a minibatch, and
$T_{\mathrm{train}}$ is the TU training-step time. Because the queue is
centralized, this lag can be estimated directly from queue state and measured
system timings.

This bounded-lag regime is compatible with GRPO~\cite{deepseekmath}: prior work
suggests that GRPO-style updates are robust to moderate off-policy sampling when
policy drift remains small~\cite{mroueh2025revisitinggrouprelativepolicy}. ASP does not rely on this robustness alone; it explicitly controls the systems sources of lag through scheduling, bounded queueing, and frequent weight refresh. 
We quantify the resulting policy lag and evaluate convergence under this lag in Section~\ref{sec:convergence}.
A complete off-policy discussion, including the update-rate / version-lag distinction and duty-cycle calculation, is given in Appendix~\ref{app:offpolicy_analysis}. The
low-level weight-flow implementation is given in Appendix~\ref{app:weightflow}.

\vspace{-0.02in}
\section{Experiments}
\label{sec:results}
In this section, we evaluate whether ASP improves the efficiency of RL post-training without destabilizing learning. We study (i) the complete-system performance against MindSpeed-RL and verl; (ii) the contribution of scheduling within a fixed ASP architecture; (iii) scalability with additional generation resources; and (iv) convergence under the measured policy lag.

\subsection{Setup}

We implement ASP inside MindSpeed-RL using GRPO~\cite{deepseekmath}. Experiments run on Ascend 910B3 nodes, each with 8 NPUs. We evaluate end-to-end performance using the MoE model \texttt{Qwen3-30B-A3B} and the dense model \texttt{Qwen3-32B}~\cite{qwen3-techreport}. Scheduling ablations first evaluate padding and load-balance metrics using fixed measured responses from \texttt{Qwen3-32B} and \texttt{Qwen3-8B}. We then execute real \texttt{Qwen3-8B} generation with simulated training to isolate the impact of scheduling on generation efficiency, iteration time, and queue behavior.
We compare ASP against two baselines. MindSpeed-RL naturally serves as our hybrid baseline, as ASP is built on top of its infrastructure; it does not decouple rollout and training into a bounded-lag multi-instance pipeline. 
We also evaluate a verl-based asynchronous and disaggregated baseline~\cite{verl}. 
This baseline evaluates the benefit of ASP beyond rollout/training separation: both verl and ASP decouple rollout and training, but only ASP applies the offline scheduling pipeline from Section~\ref{sec:scheduling} to make multi-instance rollout locally efficient and globally balanced over time.

We use two datasets. \textsc{Deepscaler}~\cite{deepscaler2025} contains challenging mathematical problems and is used for throughput, scheduling, and scalability experiments. With an 8192-token output cap on this dataset, the output-length predictors for \texttt{Qwen3-30B-A3B}, \texttt{Qwen3-32B}, and \texttt{Qwen3-8B} reach Kendall's $\tau=0.64$, $0.44$, and $0.58$, respectively. \textsc{Maze-Reasoning-GRPO-v0.1}~\cite{dao2025alphamazeenhancinglargelanguage}
is used to evaluate convergence on a task less likely to be saturated by the
base pre-trained model. We report two maze rewards: \emph{maze integrity}, measuring whether the generated solution is well formed, and \emph{maze correctness}, measuring
whether the decoded path solves the maze. Additional dataset details,
hyperparameters, node allocations, output-length distributions, and predictor
evaluation metrics are deferred to Appendix~\ref{app:experimental_details}.

To evaluate system efficiency, we measure \emph{end-to-end iteration time}, including rollout acquisition, reward/scoring, log-probability computation, policy update, and asynchronous delays visible to the iteration. We define \emph{end-to-end tokens per second} (TPS) as the number of generated tokens per iteration divided by this time and normalized by the total number of NPUs. Mean iteration times used for performance comparisons are arithmetic averages over all measured iterations, including startup.


\subsection{End-to-end performance}



Table~\ref{tab:asp-vs-baseline} compares complete ASP with MindSpeed-RL and verl using the arithmetic mean of end-to-end iteration times. 
Across both models, ASP reduces mean iteration time by 59.9--62.5\% relative to verl and by 75.1--76.8\% relative to MindSpeed-RL. For \texttt{Qwen3-30B-A3B}, complete ASP also reaches 58 tokens/s/NPU compared with 18 tokens/s/NPU for MindSpeed-RL, yielding a $3.22\times$ end-to-end throughput improvement.

\begin{table}[h!]
  \centering
    \caption{Mean end-to-end iteration time under matched eight-node resources. Reductions are relative to each baseline; lower is better. Configuration details are given in Appendix~\ref{app:run_configurations}.}
  \label{tab:asp-vs-baseline}
  \small
  \setlength{\tabcolsep}{4pt}
  \begin{tabular}{lccccc}
    \toprule
    \textbf{Model}
      & \textbf{ASP}
      & \textbf{verl}
      & \textbf{MindSpeed-RL}
      & \multicolumn{2}{c}{\textbf{ASP reduction}} \\
    \cmidrule(lr){5-6}
      & \multicolumn{3}{c}{\textbf{Mean iteration time (s)}}
      & \textbf{vs. verl}
      & \textbf{vs. MindSpeed-RL} \\
    \midrule
    \texttt{Qwen3-30B-A3B}
      & \textbf{215.2}
      & 536.7
      & 927
      & 59.9\%
      & 76.8\% \\
    \texttt{Qwen3-32B}
      & \textbf{230.7}
      & 615.1
      & 928
      & 62.5\%
      & 75.1\% \\
    \bottomrule
  \end{tabular}
\end{table}



These comparisons measure the complete systems rather than the scheduler in
isolation: ASP and the baselines also differ in execution architecture and
implementation. Section~\ref{sec:scheduling_ablation} therefore holds the ASP
architecture fixed and separately attributes the gains due to scheduling.

\subsection{Scheduling ablation}
\label{sec:scheduling_ablation}

We measure the scheduling pipeline's performance by isolating it within a fixed ASP architecture through five scenarios (C0--C4). C0 uses naïve fixed-size minibatches in the original order; C1 sorts prompts by predicted output length, exposing sorting bubbles; C2 adds dynamic programming to refine minibatch boundaries according to the second step of ASP; C3 randomly shuffles C2 minibatches to test whether randomization alone is sufficient to address sorting bubbles; and C4 applies ASP's complete balanced ordering to measure the additional benefit of explicit load balancing.
\vspace{-0.05in}
\paragraph{Local efficiency and dispatch balance.}
Table~\ref{tab:staged-scheduling}(a) evaluates C0--C4 on 39{,}000 measured
\texttt{Qwen3-32B} responses. We construct noisy length predictions calibrated to
Kendall correlation $\tau=0.64$; this is a sensitivity-test condition,
not the measured accuracy of the 32B predictor. 
Metrics use measured response lengths and are averaged over three seeds. Intra-minibatch dispersion is the standard deviation (STD) of response lengths within each minibatch, averaged over minibatches. Mean padding rate is obtained by computing
$1 - \frac{\text{mean output length}}{\text{maximum output length}}$
for each minibatch and taking their average using true output lengths. Dispatch imbalance is measured by the coefficient of variation (CV), that is the standard deviation divided by the mean, of total response tokens per consecutive batch group. Group-load spread is defined as the ratio of the largest to the smallest batch-group load. Within-minibatch dispersion and padding rate characterize the local
length heterogeneity underlying padding and skewness bubbles; group-load
CV characterizes the temporal imbalance underlying sorting bubbles.

\begin{table}[h!]
  \centering
  
  \caption{Ablation of scheduling steps. Panel (a) measures schedule properties using fixed response lengths; panel (b) executes real generation with simulated training-step durations and weight-refresh delays. Trend is the Spearman correlation between iteration index and iteration time; values near zero indicate little monotonic drift.}
  \label{tab:staged-scheduling}
  \small
  \setlength{\tabcolsep}{5pt}
  \begin{tabular}{@{}lrrrrr@{}}
    \toprule
    \textbf{Metric} & \textbf{C0} & \textbf{C1} & \textbf{C2}
      & \textbf{C3} & \textbf{C4} \\
    \midrule
    \multicolumn{6}{@{}l}{\emph{(a) Qwen3-32B responses;
      controlled $\tau=0.64$; three-seed means}} \\
    Intra-minibatch length STD (tokens)
      & 2138 & 1163 & 1163 & 1163 & 1163 \\
    Mean padding rate
      & 0.322 & 0.226 & 0.224 & 0.224 & 0.224 \\
    Batch-group load CV
      & 0.103 & 0.403 & 0.401 & 0.162 & 0.070 \\
    Group-load spread (max/min)
      & 1.68 & 9.88 & 9.81 & 2.39 & 1.39 \\
    \addlinespace
    \multicolumn{6}{@{}l}{\emph{(b) Qwen3-8B generation with simulated training;
      measured predictor $\tau=0.58$}} \\
    Mean iteration time (s)
      & 188 & 185 & 183 & 190 & 178 \\
    Iteration-time trend
      & $-0.080$ & $+0.670$ & $+0.510$ & $+0.190$ & $-0.075$ \\
    Training Queue STD (samples)
      & 41 & 45 & 40 & 36 & 36 \\
    Training Queue maximum (samples)
      & 192 & 224 & 176 & 104 & 110 \\
    \bottomrule
  \end{tabular}
\end{table}

C1 shows that sorting improves local efficiency but increases load CV from 0.103 to 0.403, resulting in substantial fluctuations in the workload dispatched over time. The dynamic programming step in C2 slightly refines padding rates (0.226 to 0.224), while leaving this temporal imbalance largely unchanged. Balanced ordering (C4) then reduces the load CV of the same minibatches from 0.401 to 0.070 without changing their padding rate, whereas random shuffling (C3) only reaches 0.162. Thus, balanced ordering preserves the local batching gains while
reducing dispatch imbalance below C0 in this condition.

\vspace{-0.1in}

\paragraph{Impact on iteration time and throughput.}
Table~\ref{tab:staged-scheduling}(b) reports results for the five schedules C0--C4 using real \texttt{Qwen3-8B} generation through vLLM actors, including ASP's dispatch, queue, and backpressure mechanisms.
To isolate scheduling effects, we simulate training-step duration and weight-refresh delay using fixed timings measured in complete ASP runs. Appendix~\ref{app:scheduling_replay} provides the details.
Sorted dispatch exhibits positive iteration-time trends (C1: $+0.670$; C2: $+0.510$), whereas C4 shows little monotonic drift ($-0.075$) and has the lowest mean time, 178\,s versus 188\,s for C0, a 5.3\% reduction.
Shuffling matches C4's queue STD and has a slightly lower queue maximum, but a higher mean iteration time (190\,s) and trend ($+0.19$).

Separately, we compare complete training runs with and without scheduling on \texttt{Qwen3-30B-A3B} (Figure~\ref{fig:schedablation}, Appendix~\ref{app:scheduling_onoff}). Enabling scheduling reduces the mean iteration time from 238 to 212\,s (10.9\%) and increases end-to-end throughput from 47 to 58 tokens/s/NPU (23.4\%).
Scheduling also reduces the STD of minibatch generation time
by 29.4\% and Training Queue STD by 15.0\%, showing that smoother generation and queue behavior also occur during complete training. First-iteration time also falls from 1042 to 724\,s (30.5\%).

\vspace{-0.1in}
\paragraph{Sensitivity to prediction quality.}
Three-seed sweeps on measured \texttt{Qwen3-32B} and
\texttt{Qwen3-8B} responses show that sorting's padding benefit
decreases as prediction quality deteriorates and becomes
marginally negative by $\tau=0.25$. Balanced ordering nevertheless
keeps group-load CV at or below C0 over the tested oracle-to-random
range. Appendix~\ref{app:scheduling_sweeps} provides the
perturbation protocol, measured predictor correlations, and
full three-seed results for both models.

\subsection{Scalability}

\begin{figure}[t]
\centering
\includegraphics[width=\linewidth]{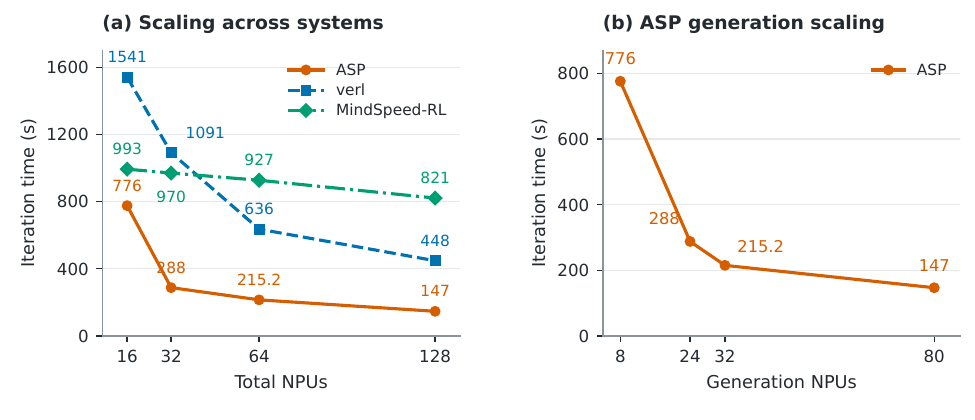}
\caption{Scalability of \texttt{Qwen3-30B-A3B}. (a) Iteration time versus total NPUs for ASP, verl, and MindSpeed-RL. (b) The same ASP results versus generation NPUs. Unlike in Table~\ref{tab:asp-vs-baseline}, verl runs without dynamic batch sizing; configurations are detailed in Appendix~\ref{app:run_configurations}.
}
\label{fig:app_scalability_comparison}
\end{figure}

We evaluate scalability using \texttt{Qwen3-30B-A3B} while varying the number of
NPUs. 
Figure~\ref{fig:app_scalability_comparison} summarizes the main results. 
As the number of NPUs increases, disaggregated multi-instance systems benefit
more than the hybrid baseline in our setting: decoupling enables phase-specific
configurations, and extra generation instances increase rollout capacity. ASP achieves lower iteration time than both baselines at every tested scale, reaching 147\,s at 128 NPUs.

\nop{
\begin{table}[htbp]
  \centering
  \caption{Scalability on Qwen3-30B-A3B. ASP benefits from additional generation
  resources until the bottleneck shifts toward training.}
  \label{tab:scalability}
  \small
  \begin{tabular}{lccc}
    \toprule
    \textbf{Total NPUs} & \textbf{ASP iter. time} & \textbf{Baseline iter. time}
    & \textbf{Reduction} \\
    \midrule
    16 & 776s & 993s & 21.85\% \\
    32 & 288s & 970s & 70.31\% \\
    64 & 216s & 927s & 76.70\% \\
    \bottomrule
  \end{tabular}
\end{table}
}

The bottleneck shifts as resources grow. At 16 NPUs, both stages are constrained: the single TU node requires recomputation and parameter offload, and the single GU node makes generation the dominant bottleneck, which explains ASP's smaller margin over the baselines. 
Adding two GU nodes to the same TU (32 NPUs) yields a nearly balanced allocation (Appendix~\ref{app:resource_allocation}) and reduces iteration time from 776\,s to 288\,s, a near-linear $2.69\times$ speedup for three times the generation resources.
At 64 NPUs, the four-node TU no longer needs recomputation or offload, and iteration time falls to 215.2\,s. At 128 NPUs, training becomes the bottleneck as it approaches its own scaling plateau, so further gains depend on training rather than rollout capacity.


\vspace{-0.1in}
\subsection{Convergence under bounded lag}
\label{sec:convergence}
Finally, we verify that asynchronous execution does not visibly harm learning in
our bounded-lag regime. On a 4-node run with 1 TU node and 3 GU nodes, the
measured timing breakdown gives, using Equation~\ref{eq:d_off},   $d_{\mathrm{off}}\approx \bar{Q}_{\mathrm{arrival}} + 3.3$, where $\bar{Q}_{\mathrm{arrival}}$ is the mean queue depth at arrival.
At this operating point, the system is rollout-bound and generated samples are
dispersed in time, so the Training Queue remains close to empty. The effective off-policy distance is therefore
approximately $3.3$ trainer steps. The full timing breakdown is reported in Appendix~\ref{app:offpolicy_analysis}.

\begin{figure}[htbp]
    \centering
    \begin{tabular}{cc}
      \includegraphics[width=0.37\linewidth]{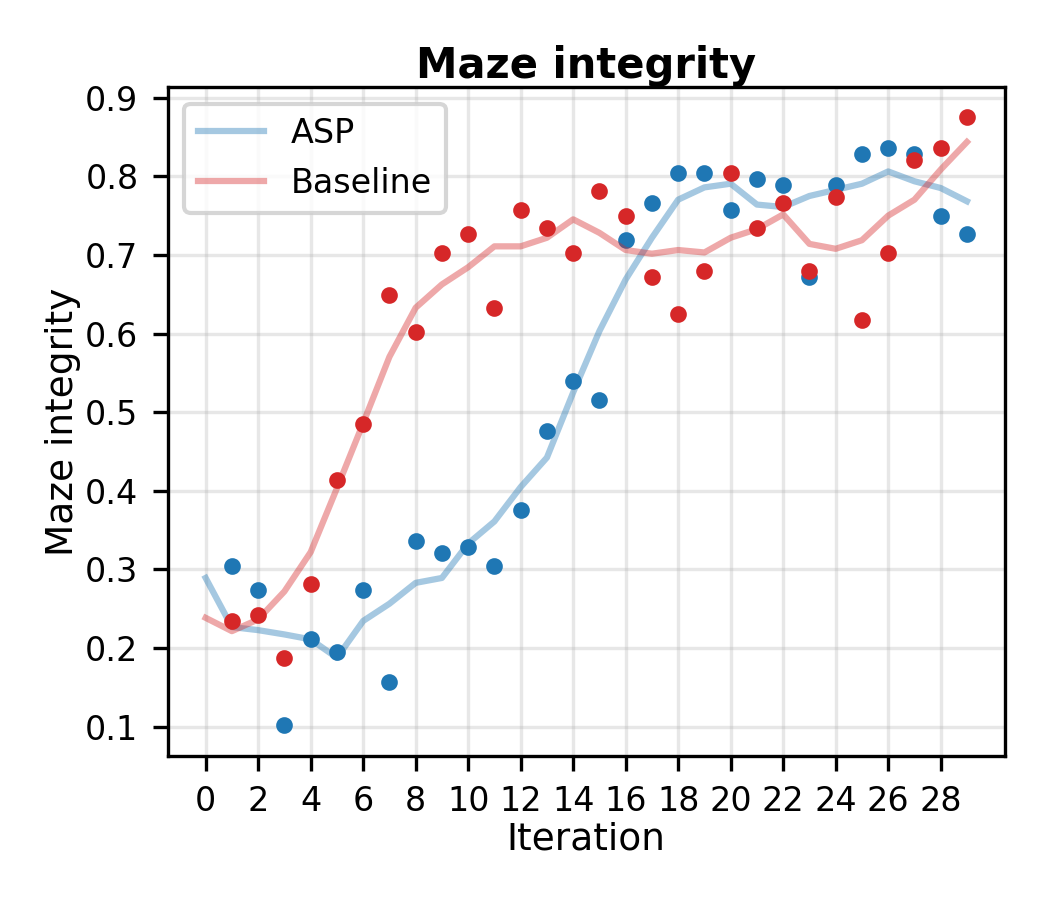} &
      \includegraphics[width=0.37\linewidth]{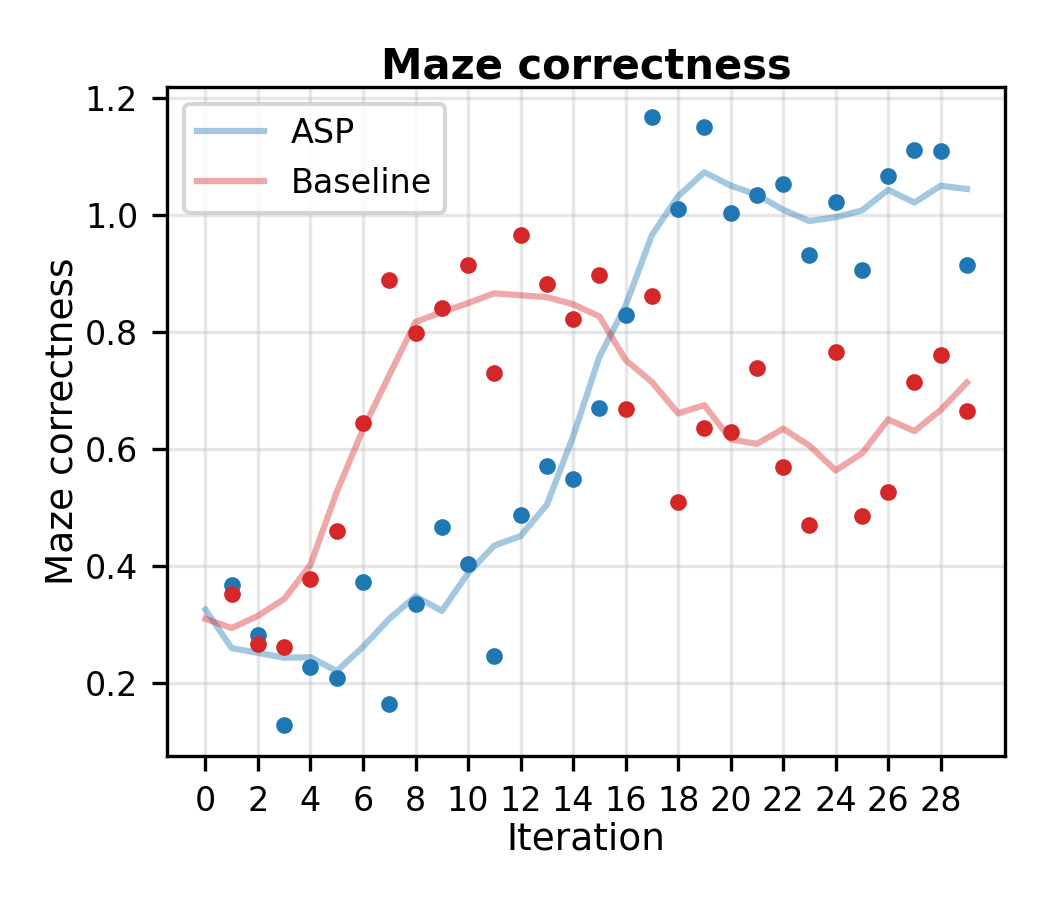} \\
      \small (a) Maze integrity reward &
      \small (b) Maze correctness reward
    \end{tabular}
    \caption{Convergence on \textsc{Maze-Reasoning-GRPO-v0.1}. ASP remains
    stable under an estimated off-policy distance of approximately 3.3 trainer
    steps.}
    \label{fig:convergence}
\end{figure}

To test whether this staleness affects learning, we run 30 GRPO iterations on \textsc{Maze-Reasoning-GRPO-v0.1} with \texttt{Qwen3-30B-A3B} and a global batch size of 16 prompts.
Figure
\ref{fig:convergence} compares ASP with MindSpeed-RL. Both methods reach
a similar maze-integrity plateau, with ASP showing a short initial delay caused
by generation warmup. Maze correctness follows the same qualitative conclusion: ASP does not show visible degradation under the measured lag. 

These results are consistent with the expected robustness of GRPO to moderate
staleness~\cite{mroueh2025revisitinggrouprelativepolicy}. More importantly,
they show that ASP's systems mechanisms---offline scheduling, bounded
queueing, and frequent weight refresh---keep the asynchronous pipeline in a
small-lag regime while delivering substantial throughput gains.


\vspace{-0.1in}
\section{Conclusion}
\label{sec:conclusion}

We presented ASP, a scheduling-centered approach for efficient asynchronous RL
post-training of LLMs. The main insight is that disaggregating rollout and
training is not sufficient by itself: multi-instance rollout must also be
scheduled so that generation is locally efficient inside each actor and globally
stable at the Training Queue level. ASP addresses this missing layer with an offline
bubble-mitigating scheduler that predicts output lengths, constructs length-homogeneous minibatches, and reorders them into balanced batch groups.

On Qwen3 MoE and dense models, ASP reduces iteration time by up to 76.8\% versus MindSpeed-RL and 62.5\% versus verl, reaching $3.22\times$ the end-to-end throughput of MindSpeed-RL. Ablations confirm scheduling's direct contribution: length-aware minibatching improves local efficiency, balanced ordering preserves these gains and stabilizes trainer arrivals, 
and complete training runs with and without scheduling show a 10.9\% reduction in iteration time within the same architecture.
Together with near-linear scaling and no visible convergence degradation at approximately 3.3 trainer steps of lag, these results demonstrate the value of coordinated minibatch construction and dispatch ordering for efficient asynchronous RL post-training.



\newpage

\begin{ack}
A significant part of this work was conducted within the Global Technology Service (GTS) team at Huawei Technologies in Dongguan, China, where Salah Chikhi and Luis Zuin were interns, and Entong Li and Li Zeng were full-time employees at the time. We thank the GTS team for its support and collaborative research environment. We also acknowledge the computing infrastructure provided by Huawei Technologies and by the Shenzhen Research Institute of Big Data.

The authors declare no competing interests.
\end{ack}

{
\small
\bibliographystyle{unsrtnat}
\bibliography{egbib}

\begin{thebibliography}{35}
\providecommand{\natexlab}[1]{#1}
\providecommand{\url}[1]{\texttt{#1}}
\expandafter\ifx\csname urlstyle\endcsname\relax
  \providecommand{\doi}[1]{doi: #1}\else
  \providecommand{\doi}{doi: \begingroup \urlstyle{rm}\Url}\fi

\bibitem[Ouyang et~al.(2022)]{ouyang2022training}
Long Ouyang et~al.
\newblock {Training language models to follow instructions with human
  feedback}.
\newblock In \emph{Advances in Neural Information Processing Systems},
  volume~35, 2022.
\newblock URL
  \url{https://proceedings.neurips.cc/paper_files/paper/2022/hash/b1efde53be364a73914f58805a001731-Abstract.html}.

\bibitem[Havrilla et~al.(2024)Havrilla, Du, Raparthy, Nalmpantis, Dwivedi-Yu,
  Zhuravinskyi, Hambro, Sukhbaatar, and
  Raileanu]{havrilla2024teachinglargelanguagemodels}
Alex Havrilla, Yuqing Du, Sharath~Chandra Raparthy, Christoforos Nalmpantis,
  Jane Dwivedi-Yu, Maksym Zhuravinskyi, Eric Hambro, Sainbayar Sukhbaatar, and
  Roberta Raileanu.
\newblock Teaching large language models to reason with reinforcement learning,
  2024.
\newblock \url{https://arxiv.org/abs/2403.04642}.

\bibitem[Wen et~al.(2025)]{wen2025rlvr}
Xuanming Wen et~al.
\newblock {Reinforcement Learning with Verifiable Rewards Implicitly
  Incentivizes Correct Reasoning in Base LLMs}, 2025.
\newblock \url{https://arxiv.org/abs/2506.14245}.

\bibitem[Hu et~al.(2025)Hu, Wu, Shen, Liu, Zhu, Wang, Jiang, Wang, Chen, Chen,
  Fang, Xianyu, Cao, Xu, and Liu]{openrlhf}
Jian Hu, Xibin Wu, Wei Shen, Jason~Klein Liu, Zilin Zhu, Weixun Wang, Songlin
  Jiang, Haoran Wang, Hao Chen, Bin Chen, Weikai Fang, Xianyu, Yu~Cao, Haotian
  Xu, and Yiming Liu.
\newblock {OpenRLHF: An Easy-to-use, Scalable and High-performance RLHF
  Framework}, 2025.
\newblock \url{https://arxiv.org/abs/2405.11143}.

\bibitem[Zhong et~al.(2025)Zhong, Zhang, Song, Hu, Jin, Wu, Chen, Chen, Zhou,
  Wan, Zhou, Jiang, Zhu, and Jiang]{streamRL}
Yinmin Zhong, Zili Zhang, Xiaoniu Song, Hanpeng Hu, Chao Jin, Bingyang Wu, Nuo
  Chen, Yukun Chen, Yu~Zhou, Changyi Wan, Hongyu Zhou, Yimin Jiang, Yibo Zhu,
  and Daxin Jiang.
\newblock {StreamRL: Scalable, Heterogeneous, and Elastic RL for LLMs with
  Disaggregated Stream Generation}, 2025.
\newblock \url{https://arxiv.org/abs/2504.15930}.

\bibitem[Han et~al.(2025)Han, You, Wang, Luo, Yang, Shi, Chen, Zhang, Lan,
  Deng, Ji, Liu, Huang, Zhang, Pan, Wang, Huang, Li, and Wu]{asyncflow}
Zhenyu Han, Ansheng You, Haibo Wang, Kui Luo, Guang Yang, Wenqi Shi, Menglong
  Chen, Sicheng Zhang, Zeshun Lan, Chunshi Deng, Huazhong Ji, Wenjie Liu,
  Yu~Huang, Yixiang Zhang, Chenyi Pan, Jing Wang, Xin Huang, Chunsheng Li, and
  Jianping Wu.
\newblock {AsyncFlow: An Asynchronous Streaming RL Framework for Efficient LLM
  Post-Training}, 2025.
\newblock \url{https://arxiv.org/abs/2507.01663}.

\bibitem[Noukhovitch et~al.(2025)Noukhovitch, Huang, Xhonneux, Hosseini,
  Agarwal, and Courville]{noukhovitch2025asynchronousrlhffasterefficient}
Michael Noukhovitch, Shengyi Huang, Sophie Xhonneux, Arian Hosseini, Rishabh
  Agarwal, and Aaron Courville.
\newblock {Asynchronous RLHF: Faster and More Efficient Off-Policy RL for
  Language Models}, 2025.
\newblock \url{https://arxiv.org/abs/2410.18252}.

\bibitem[Fu et~al.(2026)Fu, Gao, Shen, Zhu, Mei, He, Xu, Wei, Mei, Wang, Yang,
  Yuan, and Wu]{fu2026areallargescaleasynchronousreinforcement}
Wei Fu, Jiaxuan Gao, Xujie Shen, Chen Zhu, Zhiyu Mei, Chuyi He, Shusheng Xu,
  Guo Wei, Jun Mei, Jiashu Wang, Tongkai Yang, Binhang Yuan, and Yi~Wu.
\newblock {AReaL: A Large-Scale Asynchronous Reinforcement Learning System for
  Language Reasoning}, 2026.
\newblock URL \url{https://arxiv.org/abs/2505.24298}.

\bibitem[Sheng et~al.(2025{\natexlab{a}})Sheng, Tong, Wan, Zhang, Jia, Wu, Wu,
  Li, Zhang, Peng, Lin, Liu, and Wu]{sheng2025laminarscalableasynchronousrl}
Guangming Sheng, Yuxuan Tong, Borui Wan, Wang Zhang, Chaobo Jia, Xibin Wu, Yuqi
  Wu, Xiang Li, Chi Zhang, Yanghua Peng, Haibin Lin, Xin Liu, and Chuan Wu.
\newblock {Laminar: A Scalable Asynchronous RL Post-Training Framework},
  2025{\natexlab{a}}.
\newblock URL \url{https://arxiv.org/abs/2510.12633}.

\bibitem[Feng et~al.(2025)Feng, Pan, Guo, Mei, Ning, Zhang, Liu, Zhou, Shu,
  Liu, et~al.]{MindSpeed-RL}
Laingjun Feng, Chenyi Pan, Xinjie Guo, Fei Mei, Benzhe Ning, Jianxiang Zhang,
  Xinyang Liu, Beirong Zhou, Zeng Shu, Chang Liu, et~al.
\newblock {MindSpeed RL: Distributed Dataflow for Scalable and Efficient RL
  Training on Ascend NPU Cluster}.
\newblock \emph{arXiv preprint arXiv:2507.19017}, 2025.
\newblock \url{https://gitcode.com/Ascend/MindSpeed-RL}.

\bibitem[Shao et~al.(2024)Shao, Wang, Zhu, Xu, Song, Bi, Zhang, Zhang, Li, Wu,
  et~al.]{deepseekmath}
Zhihong Shao, Peiyi Wang, Qihao Zhu, Runxin Xu, Junxiao Song, Xiao Bi, Haowei
  Zhang, Mingchuan Zhang, YK~Li, Y~Wu, et~al.
\newblock {DeepSeekMath: Pushing the Limits of Mathematical Reasoning in Open
  Language Models}, 2024.
\newblock \url{https://arxiv.org/abs/2402.03300}.

\bibitem[Yang et~al.(2025)Yang, Li, Yang, Zhang, Hui, Zheng, Yu, Gao, Huang,
  Lv, Zheng, Liu, Zhou, Huang, Hu, Ge, Wei, Lin, Tang, Yang, Tu, Zhang, Yang,
  Yang, Zhou, Zhou, Lin, Dang, Bao, Yang, Yu, Deng, Li, Xue, Li, Zhang, Wang,
  Zhu, Men, Gao, Liu, Luo, Li, Tang, Yin, Ren, Wang, Zhang, Ren, Fan, Su,
  Zhang, Zhang, Wan, Liu, Wang, Cui, Zhang, Zhou, and Qiu]{qwen3-techreport}
An~Yang, Anfeng Li, Baosong Yang, Beichen Zhang, Binyuan Hui, Bo~Zheng, Bowen
  Yu, Chang Gao, Chengen Huang, Chenxu Lv, Chujie Zheng, Dayiheng Liu, Fan
  Zhou, Fei Huang, Feng Hu, Hao Ge, Haoran Wei, Huan Lin, Jialong Tang, Jian
  Yang, Jianhong Tu, Jianwei Zhang, Jianxin Yang, Jiaxi Yang, Jing Zhou,
  Jingren Zhou, Junyang Lin, Kai Dang, Keqin Bao, Kexin Yang, Le~Yu, Lianghao
  Deng, Mei Li, Mingfeng Xue, Mingze Li, Pei Zhang, Peng Wang, Qin Zhu, Rui
  Men, Ruize Gao, Shixuan Liu, Shuang Luo, Tianhao Li, Tianyi Tang, Wenbiao
  Yin, Xingzhang Ren, Xinyu Wang, Xinyu Zhang, Xuancheng Ren, Yang Fan, Yang
  Su, Yichang Zhang, Yinger Zhang, Yu~Wan, Yuqiong Liu, Zekun Wang, Zeyu Cui,
  Zhenru Zhang, Zhipeng Zhou, and Zihan Qiu.
\newblock {Qwen3 Technical Report}, 2025.
\newblock URL \url{https://arxiv.org/abs/2505.09388}.

\bibitem[Seed(2026)]{verl}
ByteDance Seed.
\newblock {verl: Volcano Engine Reinforcement Learning for LLMs}, 2026.
\newblock \url{https://github.com/verl-project/verl}.

\bibitem[Sheng et~al.(2025{\natexlab{b}})Sheng, Zhang, Ye, Wu, Zhang, Zhang,
  Peng, Lin, and Wu]{hybridflow}
Guangming Sheng, Chi Zhang, Zilingfeng Ye, Xibin Wu, Wang Zhang, Ru~Zhang,
  Yanghua Peng, Haibin Lin, and Chuan Wu.
\newblock {HybridFlow: A Flexible and Efficient RLHF Framework}.
\newblock In \emph{Proceedings of the Twentieth European Conference on Computer
  Systems}, pages 1279--1297, 2025{\natexlab{b}}.
\newblock \doi{10.1145/3689031.3696075}.
\newblock URL \url{https://doi.org/10.1145/3689031.3696075}.

\bibitem[{Qwen} et~al.(2024){Qwen}, Yang, Yang, Zhang, Hui, Zheng, Yu, Li, Liu,
  Huang, Wei, Lin, Yang, Tu, Zhang, Yang, Yang, Zhou, Lin, Dang, Lu, Bao, Yang,
  Yu, Li, Xue, Zhang, Zhu, Men, Lin, Li, Tang, Xia, Ren, Ren, Fan, Su, Zhang,
  Wan, Liu, Cui, Zhang, and Qiu]{qwen25-techreport}
{Qwen}, An~Yang, Baosong Yang, Beichen Zhang, Binyuan Hui, Bo~Zheng, Bowen Yu,
  Chengyuan Li, Dayiheng Liu, Fei Huang, Haoran Wei, Huan Lin, Jian Yang,
  Jianhong Tu, Jianwei Zhang, Jianxin Yang, Jiaxi Yang, Jingren Zhou, Junyang
  Lin, Kai Dang, Keming Lu, Keqin Bao, Kexin Yang, Le~Yu, Mei Li, Mingfeng Xue,
  Pei Zhang, Qin Zhu, Rui Men, Runji Lin, Tianhao Li, Tianyi Tang, Tingyu Xia,
  Xingzhang Ren, Xuancheng Ren, Yang Fan, Yang Su, Yichang Zhang, Yu~Wan,
  Yuqiong Liu, Zeyu Cui, Zhenru Zhang, and Zihan Qiu.
\newblock {Qwen2.5 Technical Report}, 2024.
\newblock URL \url{https://arxiv.org/abs/2412.15115}.

\bibitem[Kendall(1938)]{kendall1938rank}
M.~G. Kendall.
\newblock A new measure of rank correlation.
\newblock \emph{Biometrika}, 30\penalty0 (1-2):\penalty0 81--93, 1938.
\newblock \doi{10.2307/2332226}.
\newblock \url{https://doi.org/10.2307/2332226}.

\bibitem[Kirkpatrick et~al.(1983)Kirkpatrick, Gelatt, and
  Vecchi]{kirkpatrick1983optimization}
Scott Kirkpatrick, C.~D. Gelatt, and M.~P. Vecchi.
\newblock Optimization by simulated annealing.
\newblock \emph{Science}, 220\penalty0 (4598):\penalty0 671--680, 1983.

\bibitem[Kwon et~al.(2023)Kwon, Li, Zhuang, Sheng, Zheng, Yu, Gonzalez, Zhang,
  and Stoica]{kwon2023efficientmemorymanagementlarge}
Woosuk Kwon, Zhuohan Li, Siyuan Zhuang, Ying Sheng, Lianmin Zheng, Cody~Hao Yu,
  Joseph~E. Gonzalez, Hao Zhang, and Ion Stoica.
\newblock {Efficient Memory Management for Large Language Model Serving with
  PagedAttention}.
\newblock In \emph{Proceedings of the 29th Symposium on Operating Systems
  Principles}, pages 611--626, 2023.
\newblock \doi{10.1145/3600006.3613165}.
\newblock URL \url{https://doi.org/10.1145/3600006.3613165}.

\bibitem[Mroueh et~al.(2025)Mroueh, Dupuis, Belgodere, Nitsure, Rigotti,
  Greenewald, Navratil, Ross, and
  Rios]{mroueh2025revisitinggrouprelativepolicy}
Youssef Mroueh, Nicolas Dupuis, Brian Belgodere, Apoorva Nitsure, Mattia
  Rigotti, Kristjan Greenewald, Jiri Navratil, Jerret Ross, and Jesus Rios.
\newblock {Revisiting Group Relative Policy Optimization: Insights into
  On-Policy and Off-Policy Training}, 2025.
\newblock \url{https://arxiv.org/abs/2505.22257}.

\bibitem[Luo et~al.(2025)Luo, Tan, Wong, Shi, Tang, Roongta, Cai, Luo, Li,
  Popa, and Stoica]{deepscaler2025}
Michael Luo, Sijun Tan, Justin Wong, Xiaoxiang Shi, William~Y. Tang, Manan
  Roongta, Colin Cai, Jeffrey Luo, Li~Erran Li, Raluca~Ada Popa, and Ion
  Stoica.
\newblock {DeepScaleR: Surpassing O1-Preview with a 1.5B Model by Scaling RL}.
\newblock
  \href{https://pretty-radio-b75.notion.site/DeepScaleR-Surpassing-O1-Preview-with-a-1-5B-Model-by-Scaling-RL-19681902c1468005bed8ca303013a4e2}{Notion
  blog post}, 2025.

\bibitem[Dao and Vu(2025)]{dao2025alphamazeenhancinglargelanguage}
Alan Dao and Dinh~Bach Vu.
\newblock {AlphaMaze: Enhancing Large Language Models' Spatial Intelligence via
  GRPO}, 2025.
\newblock \url{https://arxiv.org/abs/2502.14669}.

\bibitem[Schulman et~al.(2017)]{schulman2017ppo}
John Schulman et~al.
\newblock {Proximal Policy Optimization Algorithms}, 2017.
\newblock URL \url{https://arxiv.org/abs/1707.06347}.

\bibitem[Zhao et~al.(2023)Zhao, Gu, Varma, Luo, Huang, Xu, Wright, Shojanazeri,
  Ott, Shleifer, Desmaison, Balioglu, Damania, Nguyen, Chauhan, Hao, and
  Li]{zhao2023pytorchfsdp}
Yanli Zhao, Andrew Gu, Rohan Varma, Liang Luo, Chien-Chin Huang, Min Xu, Less
  Wright, Hamid Shojanazeri, Myle Ott, Sam Shleifer, Alban Desmaison, Can
  Balioglu, Pritam Damania, Bernard Nguyen, Geeta Chauhan, Yuchen Hao, and Shen
  Li.
\newblock {PyTorch FSDP: Experiences on Scaling Fully Sharded Data Parallel}.
\newblock \emph{Proceedings of the VLDB Endowment}, 16\penalty0 (12):\penalty0
  3848--3860, 2023.
\newblock \url{https://arxiv.org/abs/2304.11277}.

\bibitem[Shoeybi et~al.(2019)]{shoeybi2019megatronlm}
Mohammad Shoeybi et~al.
\newblock {Megatron-LM: Training Multi-Billion Parameter Language Models Using
  Model Parallelism}.
\newblock \emph{arXiv:1909.08053}, 2019.

\bibitem[Zheng et~al.(2024)Zheng, Yin, Xie, Sun, Huang, Yu, Cao, Kozyrakis,
  Stoica, Gonzalez, Barrett, and Sheng]{zheng2023sglang}
Lianmin Zheng, Liangsheng Yin, Zhiqiang Xie, Chuyue Sun, Jeff Huang, Cody~Hao
  Yu, Shiyi Cao, Christos Kozyrakis, Ion Stoica, Joseph~E. Gonzalez, Clark
  Barrett, and Ying Sheng.
\newblock {SGLang: Efficient Execution of Structured Language Model Programs}.
\newblock In \emph{Advances in Neural Information Processing Systems},
  volume~37, pages 62557--62583, 2024.
\newblock URL
  \url{https://proceedings.neurips.cc/paper_files/paper/2024/hash/724be4472168f31ba1c9ac630f15dec8-Abstract-Conference.html}.

\bibitem[Moritz et~al.(2018)]{ray}
Philipp Moritz et~al.
\newblock {Ray: A Distributed Framework for Emerging AI Applications}.
\newblock In \emph{OSDI}, 2018.
\newblock \url{https://arxiv.org/pdf/1712.05889}.

\bibitem[{Microsoft DeepSpeed Team}(2020)]{deepspeed}
{Microsoft DeepSpeed Team}.
\newblock {DeepSpeed: Extreme-scale model training for everyone}, 2020.
\newblock
  \url{https://www.microsoft.com/en-us/research/blog/deepspeed-extreme-scale-model-training-for-everyone/}.

\bibitem[Wolf et~al.(2020)]{wolf2020transformers}
Thomas Wolf et~al.
\newblock {Transformers: State-of-the-Art Natural Language Processing}.
\newblock In \emph{EMNLP: System Demonstrations}, 2020.
\newblock \url{https://arxiv.org/abs/1910.03771}.

\bibitem[Rafailov et~al.(2023)Rafailov, Sharma, Mitchell, Manning, Ermon, and
  Finn]{rafailov2023dpo}
Rafael Rafailov, Archit Sharma, Eric Mitchell, Christopher~D. Manning, Stefano
  Ermon, and Chelsea Finn.
\newblock {Direct Preference Optimization: Your Language Model is Secretly a
  Reward Model}.
\newblock In \emph{Advances in Neural Information Processing Systems},
  volume~36, 2023.
\newblock URL
  \url{https://proceedings.neurips.cc/paper_files/paper/2023/hash/a85b405ed65c6477a4fe8302b5e06ce7-Abstract-Conference.html}.

\bibitem[{THUDM}(2025)]{slime}
{THUDM}.
\newblock {slime: An LLM Post-Training Framework for RL Scaling}, 2025.
\newblock \url{https://github.com/THUDM/slime}.

\bibitem[Espeholt et~al.(2018)]{impala}
Lasse Espeholt et~al.
\newblock {IMPALA: Importance Weighted Actor-Learner Architectures}.
\newblock In \emph{ICML}, 2018.
\newblock \url{https://arxiv.org/abs/1802.01561}.

\bibitem[Huang et~al.(2023)Huang, Weng, Charakorn, Lin, Xu, and
  Ontañón]{cleanba}
Shengyi Huang, Jiayi Weng, Rujikorn Charakorn, Min Lin, Zhongwen Xu, and
  Santiago Ontañón.
\newblock {CLEANBA: A Reproducible and Efficient Distributed Reinforcement
  Learning Platform}, 2023.
\newblock \url{https://arxiv.org/abs/2310.00036}.

\bibitem[Zheng et~al.(2025)Zheng, Zhao, and
  Chen]{zheng2025prosperitycollapsefaroffpolicy}
Haizhong Zheng, Jiawei Zhao, and Beidi Chen.
\newblock {Prosperity before Collapse: How Far Can Off-Policy RL Reach with
  Stale Data on LLMs?}, 2025.
\newblock \url{https://arxiv.org/abs/2510.01161}.

\bibitem[{Python Software Foundation}(2024)]{pythonThreadPoolExecutor}
{Python Software Foundation}.
\newblock \emph{concurrent.futures --- Launching parallel tasks}, 2024.
\newblock Available at
  \url{https://docs.python.org/3/library/concurrent.futures.html#concurrent.futures.ThreadPoolExecutor}.

\bibitem[Metelli et~al.(2020)Metelli, Papini, Montali, and
  Restelli]{metelli2020mis}
Alberto~Maria Metelli, Matteo Papini, Nico Montali, and Marcello Restelli.
\newblock Importance sampling techniques for policy optimization.
\newblock \emph{Journal of Machine Learning Research}, 21:\penalty0 1--75,
  2020.
\newblock \url{http://jmlr.org/papers/v21/20-124.html}.

\end{thebibliography}
}



\newpage
\appendix

\section*{Appendix Overview}
This appendix provides a comprehensive and self-contained presentation of the algorithmic foundations, implementation details, in-depth analysis, and additional experimental results of our method.
Appendix \ref{app:extended_related_work} provides a more detailed discussion of related work.
In Appendix \ref{app:system_details}, the system architecture is further explained.
Appendix \ref{app:weightflow} provides the low-level implementation details behind the weight-refresh path summarized in Section~\ref{sec:integration}. 
The off-policy analysis and scheduling algorithm details are presented in Appendices~\ref{app:offpolicy_analysis} and~\ref{app:scheduling_details}, respectively.
Appendix \ref{app:experimental_details} includes additional experimental results, benchmark descriptions, and implementation settings.
Finally, several limitations and future directions are discussed in Appendix \ref{app:limitations}.
Overall, the appendix complements the main paper with additional justification,
implementation details, and experimental context.

\section{Extended related work}
\label{app:extended_related_work}

This appendix expands the comparison in Section~\ref{sec:background}. The main
paper focuses on ASP's scheduling-centered contribution; here we give additional
context on hybrid pipelines, asynchronous/disaggregated RL systems, and
off-policy actor--learner training. Table~\ref{tab:extended-system-comparison} summarizes the main differences between the pipelines related to ASP.

\subsection{Hybrid and asynchronous RL pipelines}

Large-scale RL post-training systems are commonly organized either as
\emph{hybrid} pipelines or as \emph{asynchronous/disaggregated} pipelines. In a
hybrid pipeline, the same accelerator pool alternates between rollout,
scoring/log-probability computation, and training. This design is simple and
keeps rollouts close to the current policy, but it creates idle periods and
forces generation and training to share one resource configuration. This is
often suboptimal because autoregressive generation and backpropagation have
different system profiles: generation is largely memory-bandwidth bound and
sensitive to output-length variability, whereas training is dominated by
synchronized backpropagation and often benefits from different parallelism
layouts.

Asynchronous pipelines decouple rollout from training so that generation can
continue while the trainer updates the model. This improves overlap and allows
the two stages to use different resources. Yet separation alone does not
overcome the early scaling plateau of single-instance rollout. Scaling rollout
further requires multiple generation instances, which introduces a scheduling
problem: the system must maintain high local utilization inside each inference
minibatch while keeping the global producer--consumer dynamics at the Training
Queue stable. ASP targets this missing scheduling layer.


\subsection{StreamRL}

StreamRL~\cite{streamRL} is closely related to ASP because it studies
disaggregated RL training for LLMs and also identifies rollout as a major
bottleneck. It introduces \emph{stream generation} to overlap generation and
training, and uses an output-length ranker to mitigate \emph{skewness bubbles}:
underutilization caused by long-tail output lengths. Indeed, since decoding benefits
from maintaining a large active batch, utilization drops when most sequences in a minibatch finish early and only a few long generations remain. StreamRL
addresses this by routing the top $\alpha$\% of prompts predicted to generate
long outputs to a dedicated subset of machines. It then applies longest-processing-time-first scheduling within each instance.

ASP differs from StreamRL in three main ways. First, StreamRL uses length
prediction primarily for inter-instance routing: long-tail prompts are separated
from the rest and routed to specific nodes, but the construction of each inference minibatch is not optimized
with ASP's explicit padding objective. ASP instead uses predicted lengths to
construct length-homogeneous minibatches, directly targeting intra-minibatch
padding waste. Second, StreamRL's routing depends on a static hyperparameter $\alpha$ that may be hard to tune, whereas ASP formulates minibatch construction and batch-group
ordering as optimization problems over the full prompt set. Third, ASP
identifies and targets \emph{sorting bubbles}: queue-level instability caused by
naive length-sorted dispatch. Sorting prompts by predicted length improves local
minibatch efficiency, but it can create temporal imbalance by sending mostly
short minibatches first and mostly long minibatches later. ASP separates these
goals: it first builds length-homogeneous minibatches, then reorders them into
balanced batch groups to smooth the generation rate seen by the Training Queue.

Thus, ASP is complementary to StreamRL's motivation. StreamRL demonstrates the
value of asynchronous rollout and length-aware routing; ASP studies how length
information can be used as a scheduling signal to jointly reduce
padding, skewness, and queue instability.

\subsection{AsyncFlow}

AsyncFlow~\cite{asyncflow} studies asynchronous RL execution through a
distributed data-management module, TransferQueue, which overlaps multiple RL
stages such as actor rollout, reference computation, reward computation, and
actor update. Its main contribution is a flexible system that allows different RL stages to run concurrently.

ASP targets a narrower but complementary problem. Instead of pipelining all RL
stages, ASP keeps a single controlled Training Queue between generation and
training, and focuses on scheduling rollout work so that multi-instance
generation is both locally efficient and globally stable. This choice is
motivated by our workload: inside the Training Unit, backpropagation dominates
the stage time, while the main end-to-end bottleneck is rollout. Therefore,
trainer-side pipelining would provide limited benefit compared with improving
rollout throughput, overlap, and queue stability.

This also clarifies the difference between ASP's Training Queue and a more
general distributed data-management substrate. ASP's queue is deliberately a
narrow coordination interface between generation and training. It is co-located
with the Training Unit, exposes queue occupancy for backpressure, and makes the queue contribution to policy lag directly observable. A more general
multi-stage queueing substrate provides broader flexibility, but ASP chooses
the narrower design because it is sufficient for the scheduling problem studied in this paper.

\subsection{AReaL}

AReaL~\citep{fu2026areallargescaleasynchronousreinforcement}
is a fully asynchronous RL system that decouples generation from
training, allowing the trainer to update the policy as completed
trajectories become available. To limit off-policy effects, AReaL
controls the admission rate of new generation requests according
to a maximum permitted staleness. It also supports interruptible
generation: when new policy weights arrive, ongoing decoding
is interrupted, the old KV caches are discarded and recomputed
under the updated weights, and unfinished trajectories resume.
This avoids waiting for the longest generations before refreshing
the policy, but can produce trajectories spanning multiple policy
versions. AReaL addresses this algorithmic challenge through
decoupled PPO, which separates the behavior and proximal policies.

ASP targets a different source of inefficiency. Rather than
interrupting decoding, ASP lets each generation actor complete
its current minibatch before refreshing weights, avoiding
KV-cache recomputation and mixed-version trajectories.
More importantly, ASP proactively constructs length-homogeneous
decoding minibatches and balances consecutive batch groups,
reducing local padding and skewness without inducing unstable
trainer arrivals. While AReaL manages asynchronous execution
and staleness, it does not explicitly optimize this joint
minibatch-composition and dispatch-ordering objective.
ASP operates with unmodified GRPO in its measured small-lag
regime, supported by its offline scheduling, bounded
queueing, and balanced TU/GU resource allocation, without
claiming the larger-lag robustness studied by AReaL.

\subsection{Laminar}

Laminar~\citep{sheng2025laminarscalableasynchronousrl}
addresses long-tail generation inefficiency through
trajectory-level asynchrony, allowing trajectories to be
generated and consumed independently. It introduces
hierarchical relay workers that distribute model weights
asynchronously, avoiding global synchronization stalls
between training and generation. To mitigate underutilization
when only a few long trajectories remain, Laminar further
proposes a dynamic repacking mechanism: it monitors KV-cache
utilization to identify underutilized rollout instances and
consolidates their unfinished trajectories onto fewer
instances using the same policy version. This frees other
instances to refresh their weights and generate new
trajectories, improving generation throughput.

Although both Laminar and ASP target inefficiencies caused
by heterogeneous output lengths, they operate at different
stages. Laminar reactively redistributes ongoing trajectories
after underutilization emerges, whereas ASP proactively
organizes prompts before generation using predicted output
lengths. ASP first constructs length-homogeneous minibatches
to reduce padding and skewness, then reorders them into
balanced batch groups to prevent the sorting bubbles
created by naive length sorting. Laminar does not explicitly
optimize this initial minibatch composition or the temporal
balance of trainer arrivals. Conversely, ASP does not migrate
in-flight trajectories, leaving reactive repacking as a
potentially complementary mechanism for handling residual
prediction errors. This combination remains unevaluated.

\subsection{Commonly used RL frameworks}

Beyond research systems that explicitly study asynchronous RL pipelines, ASP is
also related to widely used RL post-training frameworks. Reinforcement learning
training for LLMs usually requires a complex pipeline that combines large-scale
response generation, reward computation, policy optimization, and distributed
training. Instead of implementing each component manually, practitioners often
rely on frameworks that provide integrated support for these stages.

Many of these frameworks are actively maintained and regularly updated, making
them suitable for incorporating recent algorithmic and systems-level advances.
Consequently, they have become widely used in both academic research and
industrial LLM post-training. In this section, we briefly introduce several
representative RL frameworks.

\paragraph{verl.}
verl~\cite{verl} is an open-source RL post-training framework for large language models based on the HybridFlow design~\cite{hybridflow}. 
It provides abstractions for expressing RL training workflows that involve multiple model roles, such as actors, critics, reference models, reward models, and rollout engines. 
Its design separates high-level control logic from distributed computation, making it easier to construct complex RL dataflows such as PPO~\cite{schulman2017ppo} and GRPO~\cite{deepseekmath}. 
verl also integrates with common large-scale training and inference backends, including FSDP~\cite{zhao2023pytorchfsdp}, Megatron-LM~\cite{shoeybi2019megatronlm}, vLLM~\cite{kwon2023efficientmemorymanagementlarge}, and SGLang~\cite{zheng2023sglang}.

In the matched-resource comparison of Table~\ref{tab:asp-vs-baseline}, verl uses training-side remove-padding and dynamic batch sizing for actor/reference passes, together with vLLM continuous rollout batching.
Those optimizations do not themselves select which
prompts form an externally returned decoding minibatch or balance the
aggregate load of consecutive batch groups on the generation side. ASP's
additional contribution resides in predicted-length composition and balanced batch-group ordering for generation efficiency and stability.

\paragraph{OpenRLHF.}
OpenRLHF~\cite{openrlhf} is an open-source framework designed to simplify large scale RLHF and RL-based post-training of language models. 
It builds on commonly used distributed training and inference systems, including Ray~\cite{ray}, vLLM~\cite{kwon2023efficientmemorymanagementlarge}, DeepSpeed~\cite{deepspeed}, and Hugging Face Transformers~\cite{wolf2020transformers}.
The framework provides implementations of several alignment and RL algorithms, such as PPO~\cite{schulman2017ppo}, DPO~\cite{rafailov2023dpo}, and related policy-gradient variants.
Its main advantage is ease of use: it offers a relatively simple interface for launching distributed RLHF experiments while still supporting large models and efficient rollout generation. 
This makes OpenRLHF a practical choice for researchers and practitioners who want to prototype or scale post-training methods without building the full infrastructure from scratch.

\paragraph{slime.}
slime~\cite{slime} is an LLM post-training framework focused on reinforcement learning at scale. 
Its core design connects Megatron-LM~\cite{shoeybi2019megatronlm} for distributed model training with SGLang~\cite{zheng2023sglang} for efficient rollout generation.
This combination allows the framework to support high-throughput generation while preserving compatibility with large-scale training strategies such as tensor, pipeline, and data parallelism. 
In addition to standard RL training workflows, slime emphasizes flexible data-generation pipelines, which can be useful for agentic, reasoning-oriented, or task-specific RL settings. 
Compared with more general-purpose RLHF frameworks, slime is particularly oriented toward large-scale RL systems where tight integration between the training backend and the inference backend is important.

\subsection{Distributed actor--learner RL and policy lag}

ASP is also related to distributed actor--learner systems such as
IMPALA~\cite{impala} and later asynchronous RL implementations such as
CleanBA~\cite{cleanba}. These systems decouple acting from learning, improving
throughput but introducing policy staleness: samples may be generated by older
behavior policies than the learner's current policy. This actor--learner lag is
typically handled through importance weighting, clipping, or by keeping the lag
small enough in practice.

LLM post-training by RL has similar concerns, but the system bottlenecks are
different. Rollouts are expensive, output lengths are heavy-tailed, and weight
refresh may require converting trainer checkpoints into inference-ready
layouts. Recent work on asynchronous RLVR reports wall-clock speedups under
moderate staleness~\cite{noukhovitch2025asynchronousrlhffasterefficient}, while
M2PO studies robustness under farther off-policy regimes by controlling the
second moment of importance weights~\cite{zheng2025prosperitycollapsefaroffpolicy}.
ASP is complementary to these algorithmic directions: it does not rely only on
algorithmic robustness, but also controls the systems sources of lag through
bounded queueing, scheduling, and frequent weight refresh.


\begin{table}[t]
  \centering
  \caption{Extended comparison of asynchronous RL systems and
  their scheduling mechanisms.}
  \label{tab:extended-system-comparison}
  \footnotesize
  \setlength{\tabcolsep}{3pt}
  \resizebox{\linewidth}{!}{%
  \begin{tabular}{lcccccc}
    \toprule
    \textbf{Feature} &
    \textbf{Hybrid RL} &
    \textbf{StreamRL} &
    \textbf{AsyncFlow} &
    \textbf{AReaL} &
    \textbf{Laminar} &
    \textbf{ASP} \\
    \midrule
    Rollout/training separation
      & No & Yes & Yes & Yes & Yes & Yes \\
    Multi-instance rollout
      & No & Yes & Config.-dep. & Yes & Yes & Yes \\
    Independent training/generation configurations
      & No & Yes & Yes & Yes & Yes & Yes \\
    Predicted-length minibatch construction
      & No & No & No & No & No & Yes \\
    Long-tail rollout mitigation
      & No & Yes & No & Yes & Yes & Yes \\
    Explicit minibatch optimization for rollout padding
      & No & No & No & No & No & Yes \\
    Balanced batch-group ordering
      & No & No & No & No & No & Yes \\
    Weight-update decoding interruption
      & No & No & No & Yes & No & No \\
    In-flight trajectory repacking
      & No & No & No & No & Yes & No \\
    \bottomrule
  \end{tabular}}
\end{table}

\section{System architecture details}
\label{app:system_details}

This appendix provides implementation details omitted from
Section~\ref{sec:integration}. 
The main paper describes the architecture at the level needed to understand ASP's scheduling and bounded-lag integration; here we summarize the concrete components used in our prototype.

\subsection{ASP at a glance}

ASP is integrated into an asynchronous RL pipeline with three
main components: a \emph{Training Unit}, one or more \emph{Generation Units}, and a \emph{Control Unit}. The TU performs reward
computation, log-probability evaluation, and policy updates. Each GU hosts
multiple generation actors, typically vLLM instances~\cite{kwon2023efficientmemorymanagementlarge},
that repeatedly decode inference minibatches. The CU coordinates dispatch,
receives completed samples, and forwards them to the Training Queue.

The architecture is designed around three invariants. First, generation and training operate continuously whenever possible, instead of alternating as in a hybrid loop. Second, the Training Queue remains bounded through backpressure. Third, weight refresh is asynchronous: actors finish
their current minibatch before loading a new snapshot, avoiding a global pause of all generation workers.

\subsection{Control Unit}

The CU is a CPU-side process co-located with the TU. It is the coordination hub of the system: it dispatches scheduled minibatches to GUs, receives completed responses, forwards samples to the Training Queue, and notifies GUs when new weight snapshots are available. Keeping the CU close to the TU makes the Training Queue state directly observable and simplifies backpressure.

The CU contains three logical controllers.

\paragraph{Generation Sender.}
The Generation Sender sends orchestration messages, prompt minibatches, and
weight-snapshot notifications to the GUs. Prompts are dispatched according to
ASP's scheduled order. In particular, the sender treats the scheduler's batch
groups as dispatch units, so that consecutive groups have approximately
balanced predicted generation load. The sender also maintains the metadata
needed to merge returned responses with their original prompts before forwarding
samples to the Training Queue.

\paragraph{Database.}
The Database stores prompts that remain to be sent, tracks which minibatches
have been issued, and records failed sends for retry. 
Since ASP performs scheduling offline, the Database does not solve a new optimization problem during training; it supplies minibatches in the precomputed order as the Generation Sender replenishes the per-GU input queues.

\paragraph{Ingestor.}
The Ingestor receives completed minibatches from GUs and passes them to the
Training Queue. Separating ingestion from the training loop keeps network and
coordination work off the critical trainer path. It also allows the CU to record
per-GU send/receive events, which are useful for measuring queue stability and
debugging generation imbalance.

\subsection{Per-GU queue targets and backpressure}

To keep generation actors fed while retaining flexibility to balance work across GUs, the CU maintains a small per-GU input-queue target, typically 5--10 minibatches in our implementation. The lower bound hides network latency and avoids starving generation actors. The upper bound limits work committed to each GU, leaving more minibatches available for dispatch to other GUs as they become available. 

The Training Queue provides the main backpressure signal. When completed
samples accumulate beyond a configured threshold, the CU signals GUs to
pause the start of new decoding minibatches, including those already in
their input queues. Generation resumes when queue occupancy falls
sufficiently. This limits the buildup of completed samples and the policy
lag associated with waiting for training. In the rollout-bound
configurations used in our experiments, the Training Queue usually remains
close to empty, but this mechanism also controls queue growth when
generation temporarily outpaces training.

\subsection{Training Unit and Training Queue}

The TU follows the MindSpeed-RL training structure: it performs reward
computation, reference/log-probability computation, and policy updates using
GRPO~\cite{deepseekmath}. Unlike systems that pipeline several trainer
sub-stages, ASP does not pipeline the internal TU computation. In our workload,
backpropagation dominates the TU stage time, and the main end-to-end bottleneck
is rollout. Therefore, trainer-side pipelining would provide limited benefit
compared with improving rollout throughput and overlap.

\begin{figure}[t]
    \centering
    \includegraphics[width=0.6\linewidth]{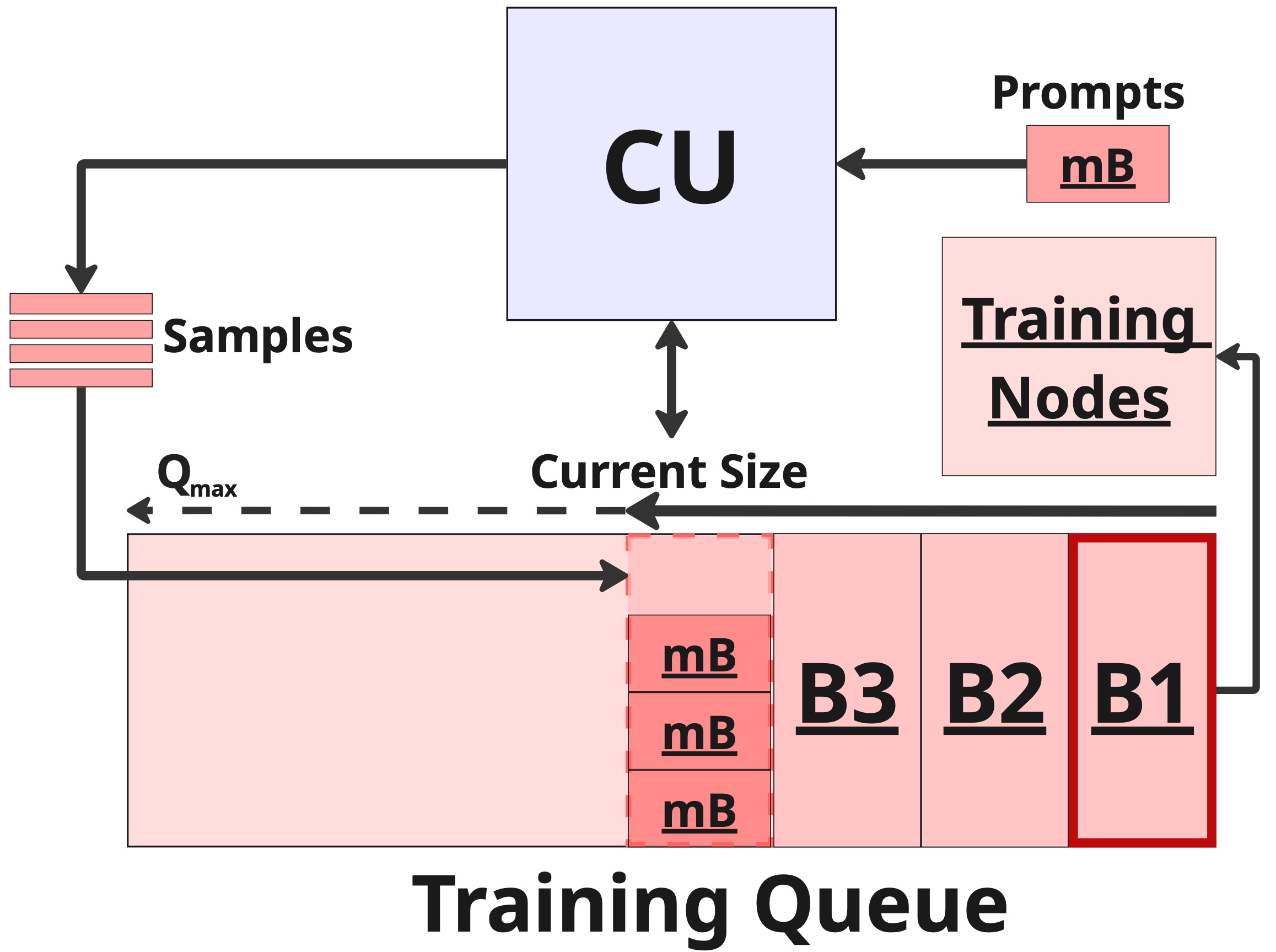}
    \caption{Training Queue. Completed inference minibatches accumulate as prompt--response samples and are consumed by the Training Unit as global training batches. mB denotes a minibatch, B a global batch, and $Q_{\rm max}$ the queue-occupancy threshold for triggering backpressure.}
    \label{fig:app_training_queue}
\end{figure}

The Training Queue is implemented as a separate Ray actor~\cite{ray}. It stores samples in a lock-protected queue and assembles them into global batches for training, as illustrated in Figure~\ref{fig:app_training_queue}. It records queue length, sample arrival events, and per-GU metadata. These counters are used both for backpressure and for estimating the effective off-policy distance.

\subsection{Generation Units and generation actors}

A GU hosts multiple generation actors. Each actor stores a copy of the inference model and repeatedly executes the same lifecycle: receive one minibatch, generate responses, return the completed samples, and then either accept another minibatch or reload weights if a newer snapshot is available. Actors are independent, so multiple minibatches can be decoded concurrently across a GU.

Each GU also has an inference-side control process that receives minibatches from the CU and assigns them to idle actors. Before assigning a new minibatch, it checks whether the CU has paused generation through its backpressure signal. While the pause is active, queued minibatches remain pending and no new decoding starts; actors already decoding finish their current minibatches.


\subsection{Resource allocation heuristic}
\label{app:resource_allocation}

ASP uses a simple static resource-allocation heuristic. The goal is to choose
the number of TU and GU nodes so that rollout throughput is close to training
consumption, while keeping generation as a mild bottleneck when possible. This
prevents unbounded Training Queue growth and keeps generation resources highly
utilized.

The procedure is:

\begin{enumerate}
    \item Determine the minimum number of nodes needed by the TU for the target
    model and global batch size, then measure the TU training-step time.

    \item Choose a generation actor configuration on the smallest convenient
    hardware slice, such as one node or one NPU group, and measure the time to
    generate one inference minibatch.

    \item Replicate generation actors across GUs until aggregate rollout
    throughput approximately matches TU consumption. We prefer the GU side to be
    a mild bottleneck, because this keeps the Training Queue bounded and avoids
    frequent pause/resume behavior.

    \item If additional nodes remain, scale the TU and repeat the balancing
    step. This alternating expansion maintains a near-balanced operating point.
\end{enumerate}

This heuristic is intentionally simple. More elaborate dynamic allocation
schemes are possible, but static allocation worked well in our setting because
ASP's scheduler reduces generation-time variance before execution and the
Training Queue absorbs short-term fluctuations online.

\section{Weight-flow implementation}
\label{app:weightflow}

This appendix provides the low-level implementation details behind the weight-refresh path summarized in Section~\ref{sec:integration}. 
The purpose of this path is to preserve independent training/generation specialization without re-coupling the two sides at every update.

\subsection{Overview}

In a disaggregated RL pipeline, generation actors must periodically receive
newer trainer weights. In ASP, this is nontrivial because the TU and GUs may use
different parallelism layouts. A trainer checkpoint therefore cannot always be
broadcast directly to inference actors. Instead, ASP uses a three-stage
weight-flow path:

\begin{enumerate}
    \item export a trainer snapshot from the TU;
    \item convert the snapshot into the layout expected by GU inference actors;
    \item reload the converted weights inside each generation actor.
\end{enumerate}

The first two stages are mostly asynchronous relative to generation. The third
stage may stall an actor, so ASP uses deferred reload: an actor finishes its
current minibatch before loading the new snapshot.

\subsection{Trainer-side export}

When the TU emits a new snapshot, each training worker prepares a CPU parameter dictionary containing its local shard in a vLLM-compatible format. This synchronous preparation is small relative to a full training iteration. The TU then launches an asynchronous task that writes these local shards to shared storage, preserving the training-side parallelism layout.

After all training nodes finish saving, the CU notifies the corresponding GU
that a new snapshot is available. In our prototype, this notification is sent
through a lightweight HTTP request containing the file-system locations of the
exported checkpoint. To bound storage growth, older snapshots are deleted
progressively; in our implementation, we retain only a small recent window.

\subsection{Generation-side conversion}

Upon notification, the GU converts the training snapshot into an inference-ready
layout. Concretely, the conversion merges training tensor-parallel ranks and
expert shards when needed, reconstructs the model expected by inference, and
repartitions it according to the GU's tensor, expert, and data parallel
configuration. This conversion is performed once per snapshot at the GU level,
rather than separately inside every generation actor.

This stage allows the TU and GUs to use different parallelism layouts. For
example, training may prefer a layout optimized for backpropagation, while
generation may prefer smaller tensor-parallel groups replicated across multiple
actors. Without conversion, the two sides would need to share a common layout,
weakening the benefit of disaggregation.

\subsection{Reload methods evaluated}

We evaluated several ways to load new weights into vLLM on Ascend NPUs. Two
straightforward approaches were too slow in our environment. First, calling
vLLM's \texttt{load\_weights} endpoint took roughly 200\,s for Qwen3-30B-A3B.
Second, restarting the entire vLLM process after each iteration had a similar
cost. These approaches were disproportionate relative to minibatch generation
time and would significantly reduce the duty cycle of generation actors.

The bottleneck came primarily from software and I/O limitations rather than raw
device bandwidth. In particular, our environment did not provide a fast
\texttt{safetensors} loading path or IPC-handle-based zero-copy sharing between
processes. As a result, repeated deserialization and serialized host-to-device
(H2D) transfers became dominant.

\subsection{Parallelized host-to-device loading}

Our most effective implementation parallelizes loading inside each vLLM worker.
We use a \texttt{ThreadPoolExecutor}~\cite{pythonThreadPoolExecutor} to drive
multiple independent device streams in parallel, each issuing H2D
\texttt{copy\_} transfers. Small tensors are grouped together to reduce launch
overhead, while large tensors are assigned to separate streams to reduce
head-of-line blocking. This improves overlap between file reads, tensor staging,
and device copies.

This logic is implemented inside vLLM's \texttt{NPUWorker}. With this design,
end-to-end reload time dropped from roughly 200\,s to about 60--75\,s for
Qwen3-30B-A3B in our prototype. The exact value depends on the model layout, file
system behavior, and concurrent load on the node. In the timing breakdown used
for our off-policy analysis, the measured reload time was
$T_{\mathrm{load}}=73.9$\,s.

\subsection{Deferred actor reload}

A naive synchronous update would ask every generation actor to stop and reload
as soon as a snapshot is available. This creates a global pause and can trigger
herd effects: many actors finish, reload, and resume at the same time,
increasing CPU and I/O pressure. ASP instead uses deferred reload. If an actor
is busy decoding a minibatch, it finishes that minibatch first. It then reloads
the latest available snapshot before accepting new work.

Deferred reload preserves execution order within each actor and avoids
interrupting in-flight generation. It adds at most one minibatch of actor-local
staleness. In practice, this tradeoff is favorable because it preserves continuous generation
and avoids a system-wide synchronization point. The corresponding generation duty-cycle estimate is given in Appendix~\ref{app:generation_duty_cycle}.



\subsection{Alternative optimizations}

Further reductions in reload overhead are possible. One practical option is to dedicate one
generation actor as a staging actor: it loads the latest converted snapshot and
then propagates weights device-to-device (D2D) to sibling actors as they finish
their current minibatch. This would replace repeated host-mediated loading with
peer-to-peer transfer and could shrink the visible reload window. Similar ideas
appear in OpenRLHF~\cite{openrlhf}.

Another optimization is to remove the shared-file-system dependency. Although
shared storage is convenient in a single-cluster prototype, it can become a
tail-latency bottleneck and is less suitable for cross-cluster deployments. A
direct TU-to-GU transfer API, or an exchange based on object storage, would reduce
disk dependence and make heterogeneous or geographically separated deployments
more practical.

\section{Off-policy analysis}
\label{app:offpolicy_analysis}

This appendix expands the bounded-lag discussion from
Section~\ref{sec:integration}. The main paper uses a compact approximation of
the off-policy distance; here we make explicit the two forms of off-policy
behavior induced by asynchronous rollout, the importance-ratio interpretation,
the multi-instance offset, and the measured timing breakdown used in our
experiments.

\subsection{Two off-policy regimes}

In asynchronous RL post-training, samples may be generated with slightly older
model weights than the weights currently used by the trainer. The relevant
question is therefore not whether the system is perfectly on-policy, but whether
this lag remains small enough to preserve stable learning. Let $\pi_t$ denote
the trainer policy after iteration $t$. We use the term
\emph{off-policy distance} $d_{\mathrm{off}}$ for the lag, measured in trainer
steps, between the behavior policy used to generate a sample and the current
trainer policy. This notion is closely related to staleness or version lag in
distributed actor--learner systems~\cite{impala,cleanba,zheng2025prosperitycollapsefaroffpolicy}.

We distinguish two regimes according to the main source of lag.

\paragraph{Type I: update-rate off-policy.}
The generation weights are refreshed once every $\Delta>1$ training iterations, so the lag increases between refreshes and drops when a newer snapshot is loaded. Larger refresh intervals therefore increase policy lag and may impair convergence.

\paragraph{Type II: version-lag off-policy.}
Even when snapshots are produced after every training iteration, snapshot export, conversion, reload, generation, and queueing introduce delay. This lag reflects the time needed for a snapshot to produce samples that reach training, rather than an intentionally reduced refresh rate. It can remain approximately constant under steady operating conditions, but varies with execution times and queue occupancy in an asynchronous pipeline.

ASP primarily targets the second regime: generation actors load the latest available snapshot after finishing their current minibatch. Training Queue backpressure limits the additional lag caused by completed samples waiting for training. Under this combination of weight refresh and queue control, the measured off-policy distance remains small, and we observe no visible convergence degradation in our experiments. This is consistent with prior work showing that off-policy GRPO can remain effective when the sampling policy stays close to the current policy~\cite{mroueh2025revisitinggrouprelativepolicy}.

\subsection{Off-policy distance}

\begin{figure}[t]
  \centering
  \includegraphics[width=0.6\linewidth]{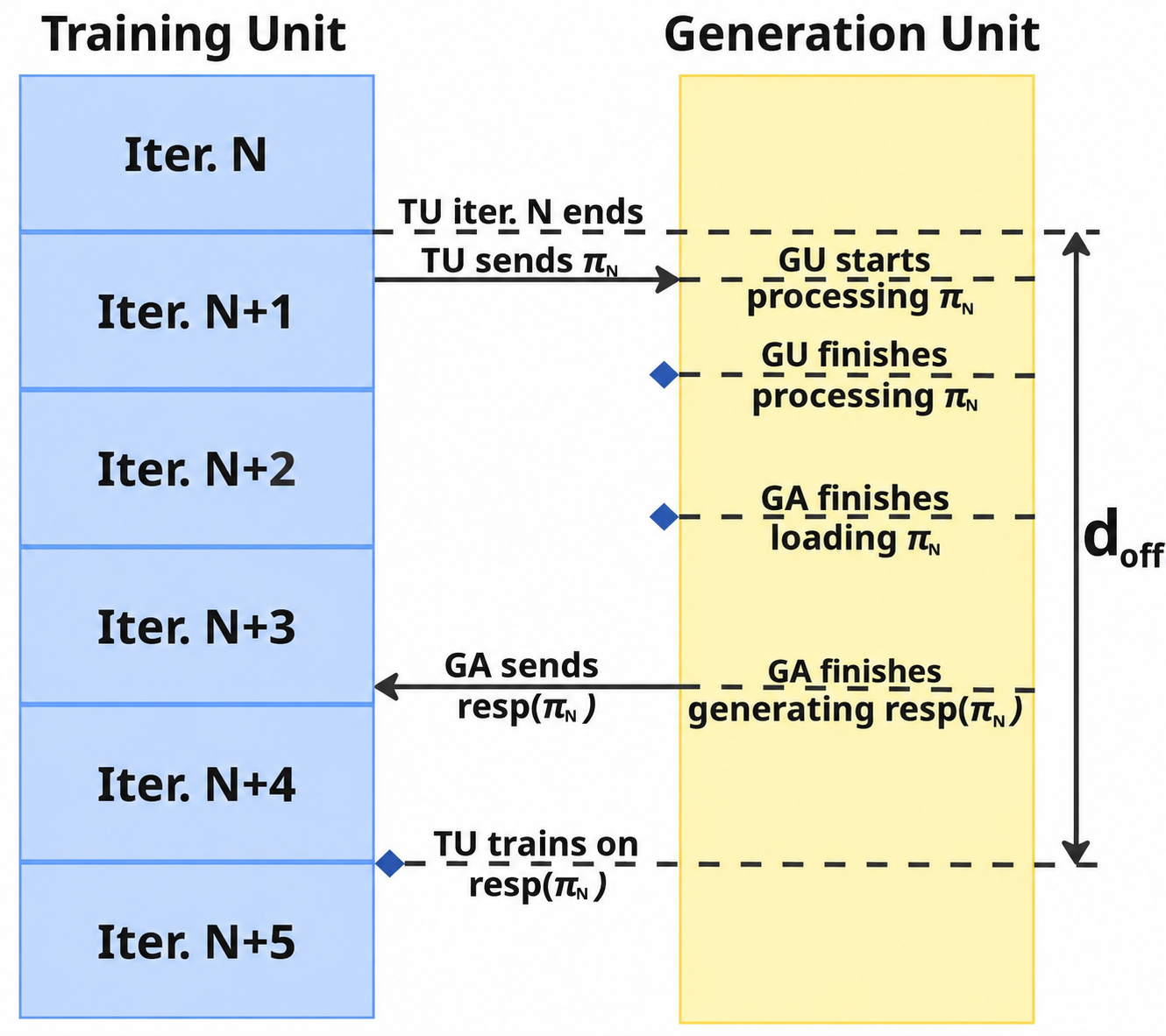}
  \caption{Off-policy distance in ASP. Samples generated with an older snapshot are consumed by a later trainer iteration; their lag, measured in trainer steps, reflects snapshot delay, generation time, and queue depth. GA denotes a generation actor.}
  \label{fig:app_distance}
\end{figure}

After the Training Unit completes iteration $t$ and emits a snapshot, the Generation Units eventually load this snapshot and produce samples with it. These samples enter the Training Queue and are consumed at a later trainer iteration, as illustrated in Figure~\ref{fig:app_distance}. The resulting distance includes snapshot export, snapshot conversion, actor reload, generation time, and queueing delay.

Operationally, for a minibatch arriving at the Training Queue, we approximate
the instantaneous off-policy distance as
\[
d_{\mathrm{off}}
\approx
Q_{\mathrm{arrival}}
+
\frac{
T_{\mathrm{save}} + T_{\mathrm{convert}} + T_{\mathrm{load}} + T_{\mathrm{gen}}
}{
T_{\mathrm{train}}
},
\]
where $Q_{\mathrm{arrival}}$ is the Training Queue depth measured in global batches immediately before the minibatch is inserted, $T_{\mathrm{save}}$ is the trainer-side
snapshot export time, $T_{\mathrm{convert}}$ is the generation-side conversion
time, $T_{\mathrm{load}}$ is the actor reload time, $T_{\mathrm{gen}}$ is the
time to generate one inference minibatch, and $T_{\mathrm{train}}$ is the
Training Unit step time.

A practical advantage of ASP's single controlled Training Queue is that the
queue contribution to this distance is directly observable at runtime. The
remaining terms are system timings that can be measured for each model and
deployment configuration. This makes the version-lag contribution easier to
monitor than in systems where coordination state is spread across several
task-specific queues or controllers.

\subsection{Importance-ratio interpretation}

In the version-lag regime, the behavior policy for a consumed sample is
$\pi_{t-d_{\mathrm{off}}}$ while the update is applied to $\pi_t$. For an
autoregressive output $y=(y_1,\ldots,y_m)$ generated from prompt $x$, the
per-token importance ratio is
\[
r_j
=
\frac{
\pi_t(y_j \mid x,y_{<j})
}{
\pi_{t-d_{\mathrm{off}}}(y_j \mid x,y_{<j})
}.
\]
The corresponding sequence-level importance weight is
\[
r(y\mid x)
=
\prod_{j=1}^{m} r_j .
\]
When the cumulative drift
$\mathrm{KL}(\pi_t \| \pi_{t-d_{\mathrm{off}}})$ is small, these ratios remain
concentrated near one. PPO-style clipping~\cite{schulman2017ppo} and
GRPO-style updates~\cite{deepseekmath} further reduce the effect of moderate
ratio deviations. ASP complements these update rules by limiting policy lag through frequent weight refresh and Training Queue backpressure.


\subsection{Multi-instance offset}

ASP uses multiple generation actors, which creates a second, smaller source of
lag variation. Different actors may briefly hold different snapshots because
they reload asynchronously: an actor that is busy decoding finishes its current
minibatch before loading the latest snapshot. As a result, samples arriving at
the Training Queue may be generated from a small mixture of nearby behavior
policies,
\[
\{\pi_{t-\delta_1}, \pi_{t-\delta_2}, \ldots\},
\]
where the $\delta_i$ values differ slightly across actors.

In practice, this actor-specific offset is small in our runs. Indeed, snapshots are
frequent, and deferred reload adds at
most one minibatch of actor-local staleness. Moreover, generated samples are
time-dispersed in the rollout-bound operating point, so the Training Queue
usually remains close to empty. We therefore did not apply additional
off-policy corrections.

Two lightweight corrections could be useful in larger-lag regimes. First,
samples could be bucketed by the snapshot that generated them, with baselines
or normalizers computed per snapshot. This would avoid mixing advantages across
slightly different behavior policies. Second, minibatches could be treated as
draws from a mixture of nearby snapshot policies, using multiple or relative
importance sampling to correct the mismatch~\cite{metelli2020mis}. We leave
these extensions for future work because the measured lag in our experiments is
already small.

\subsection{Measured timing breakdown}

The convergence experiment in Section~\ref{sec:convergence} uses a 4-node
configuration with 1 Training Unit node and 3 Generation Unit nodes. The measured
timing breakdown is
\[
T_{\mathrm{save}} = 23.5\text{s}, \quad
T_{\mathrm{convert}} = 84.6\text{s}, \quad
T_{\mathrm{load}} = 73.9\text{s}.
\]
Thus, the snapshot-related delay is
\[
T_{\mathrm{delay}}
=
T_{\mathrm{save}} + T_{\mathrm{convert}} + T_{\mathrm{load}}
=
182.0\text{s}.
\]
With average minibatch generation time
$T_{\mathrm{gen}}=763.3\text{s}$ and Training Unit step time
$T_{\mathrm{train}}=287.6\text{s}$, we obtain
\[
d_{\mathrm{off}}
\approx
\bar{Q}_{\mathrm{arrival}}
+
\frac{182.0 + 763.3}{287.6}
\approx
\bar{Q}_{\mathrm{arrival}} + 3.3 .
\]
At this operating point, the system is rollout-bound and the Training Queue
remains close to empty, so $\bar{Q}_{\mathrm{arrival}}\approx 0$ and
$d_{\mathrm{off}}\approx 3.3$ trainer steps. This is the value reported in the
main text.

\subsection{Generation duty cycle}
\label{app:generation_duty_cycle}

The same timing breakdown also estimates how much time each generation actor
spends doing useful inference. As the trainer-side save and generation-side conversion are asynchronous relative to ongoing generation, the main actor-visible stall is the reload step. Therefore, a simple per-actor duty-cycle estimate is
\[
\eta
\approx
\frac{T_{\mathrm{gen}}}{T_{\mathrm{gen}}+T_{\mathrm{load}}}.
\]
Using $T_{\mathrm{load}}=73.9\text{s}$ and $T_{\mathrm{gen}}=763.3\text{s}$ gives
\[
\eta
\approx
\frac{763.3}{763.3+73.9}
\approx
0.912 .
\]
Thus, generation actors spend about 91\% of their time performing inference in
this configuration. This supports the design choice of deferred reload: weight
refresh remains visible, but it does not dominate the generation duty cycle.

\section{Scheduling algorithm details}
\label{app:scheduling_details}

This appendix gives the full optimization objectives and pseudocode for the
offline scheduling pipeline summarized in Section~\ref{sec:scheduling}. ASP
performs three steps before training: predict output lengths, construct
length-homogeneous minibatches, and reorder minibatches into balanced batch
groups.

\subsection{Length prediction}

The scheduler uses predicted output lengths to order and group prompts. In our implementation, we collect a
small set of prompt--length pairs by running the model to be trained on prompts from the training dataset, and fine-tune Qwen~2.5 1.5B~\cite{qwen25-techreport} to predict the
corresponding output length. The predictor is kept fixed throughout training.

Ranking quality is particularly important because sorting aims to bring prompts with similar output lengths together. We report the predictor's Kendall rank correlation $\tau$~\cite{kendall1938rank} in Appendix~\ref{app:output_length_predictor} and evaluate the scheduler's sensitivity to prediction quality in Appendix~\ref{app:scheduling_sweeps}.


\subsection{Dynamic programming for minibatch construction}

\begin{figure}[t]
    \centering
    \includegraphics[width=0.6\linewidth]{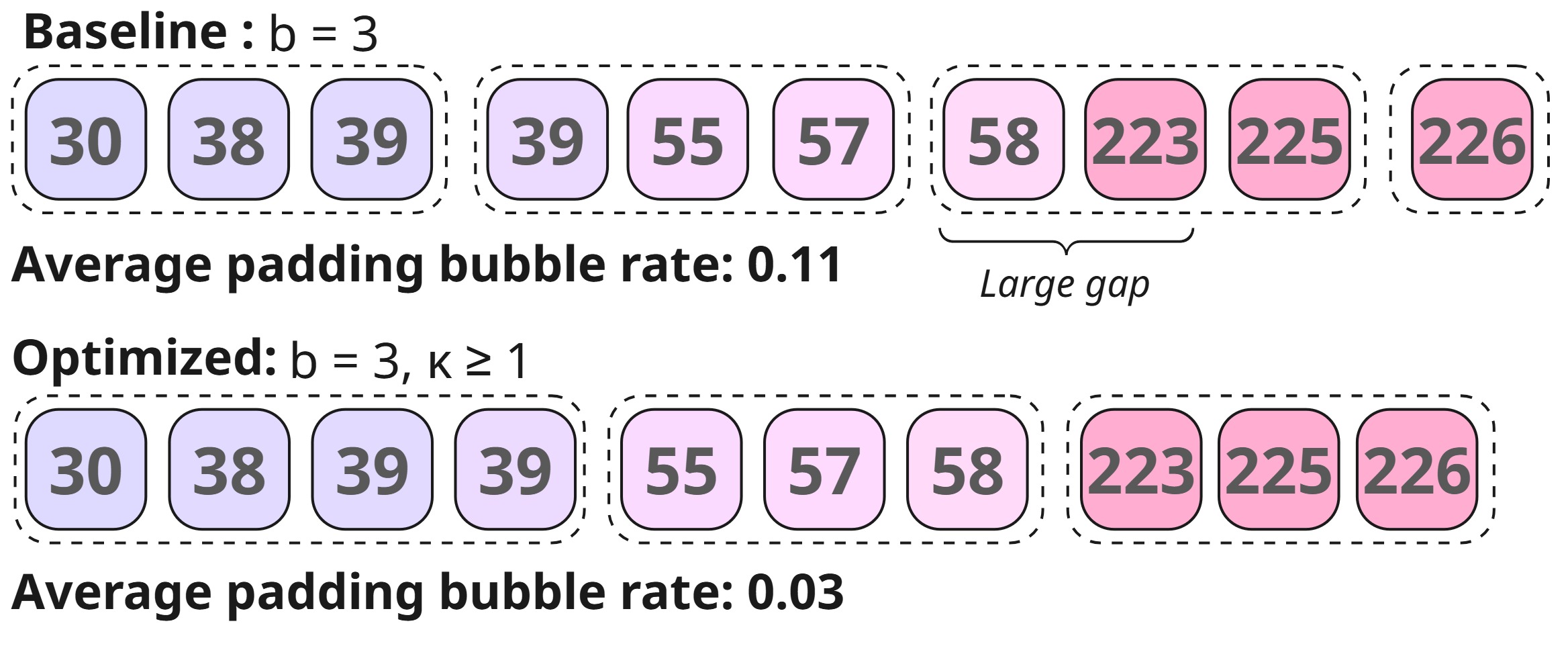}
    \caption{Minibatch construction. Allowing small variations in minibatch
    size can avoid bad boundaries in the sorted length list and reduce predicted
    padding bubbles.}
    \label{fig:app_minibatch}
\end{figure}

Let $l_0 \leq \cdots \leq l_{n-1}$ be the predicted output lengths after sorting the
dataset. The goal is to partition the sorted list into minibatches whose sizes
are between $b$ and $b+\kappa$, where $b$ is the base minibatch size and
$\kappa$ is a small allowed size variation. This flexibility allows the scheduler to adjust minibatch boundaries to reduce predicted padding, as illustrated in Figure~\ref{fig:app_minibatch}. A partition is represented by
endpoints
\[
0 = x_0 < x_1 < \cdots < x_{N_B}=n,
\]
where minibatch $i$ contains prompts with indices
$x_i,\ldots,x_{i+1}-1$. For a candidate minibatch, define
\[
R_i
=
\frac{
\sum_{j=x_i}^{x_{i+1}-1} l_j
}{
(x_{i+1}-x_i)\,l_{x_{i+1}-1}
}.
\]
This ratio compares the predicted useful generated tokens with the total token slots implied by the minibatch size and longest predicted output.
Thus, $1-R_i$
estimates the minibatch padding bubble rate. ASP aims to minimize the estimated average padding bubble rate, that is:
\[
\bar{P}
=
1
-
\frac{1}{N_B}
\sum_{i=0}^{N_B-1} R_i .
\]

ASP uses the sum of minibatch utilizations as an additive surrogate for average utilization:

\[
\begin{aligned}
& \textbf{maximize}
&&
\sum_{i=0}^{N_B-1} R_i
\\
& \textbf{subject to}
&&
0 = x_0 < x_1 < \cdots < x_{N_B}=n,
\\
&&&
b \leq x_{i+1}-x_i \leq b+\kappa
\quad
\text{for all } i.
\end{aligned}
\]

Algorithm~\ref{alg:dp} summarizes the dynamic program used to solve this additive objective. Let
$S[i]=\sum_{j=0}^{i-1} l_j$ be the prefix sum of predicted lengths. For any
candidate minibatch starting at $j$ and ending before $i$, its utilization is
\[
R(j,i)
=
\frac{S[i]-S[j]}{(i-j)l_{i-1}} .
\]
The dynamic program is
\[
\mathrm{dp}[i]
=
\max_{j \in \{\max(0,i-(b+\kappa)),\dots, i-b\}}
\left(
\mathrm{dp}[j] + R(j,i)
\right),
\]
with $\mathrm{dp}[0]=0$, unreachable states set to $-\infty$, and an empty maximization yielding $-\infty$. Backpointers recover the optimal endpoints. Since each state checks at most $\kappa+1$ candidate minibatch sizes, the complexity is
$O(\kappa n)$ time and $O(n)$ space. For fixed small $\kappa$, this is linear in
the dataset size.

\begin{algorithm}[t]
\caption{Dynamic programming for length-homogeneous minibatches}
\label{alg:dp}
\begin{algorithmic}[1]
\State \textbf{Input:} sorted predicted lengths $l[0..n-1]$, base size $b$, variation $\kappa$
\State Compute prefix sums $S[0..n]$, with $S[0]\gets 0$ and $S[i+1]\gets S[i]+l[i]$
\State Initialize $\mathrm{dp}[0]\gets 0$ and $\mathrm{dp}[i]\gets -\infty$ for $i>0$
\State Initialize $\mathrm{prev}[0..n]\gets -1$
\For{$i=1$ \textbf{to} $n$}
    \For{$j=\max(0,i-(b+\kappa))$ \textbf{to} $i-b$}
        \State $R \gets \dfrac{S[i]-S[j]}{(i-j)l[i-1]}$
        \If{$\mathrm{dp}[j]+R > \mathrm{dp}[i]$}
            \State $\mathrm{dp}[i]\gets \mathrm{dp}[j]+R$
            \State $\mathrm{prev}[i]\gets j$
        \EndIf
    \EndFor
\EndFor
\State Recover endpoints by backtracking from $n$ through $\mathrm{prev}$
\State \textbf{Output:} minibatch endpoints and predicted padding rate
\end{algorithmic}
\end{algorithm}

\subsection{Balanced batch-group ordering}

After the dynamic program, the minibatches are locally efficient but still
ordered from short to long. Dispatching them in this order would create sorting
bubbles: early groups would be fast, later groups slow, and the Training Queue
could oscillate. ASP therefore reorders minibatches into balanced batch groups.

\begin{figure}[h!]
    \centering
    \includegraphics[width=0.72\linewidth]{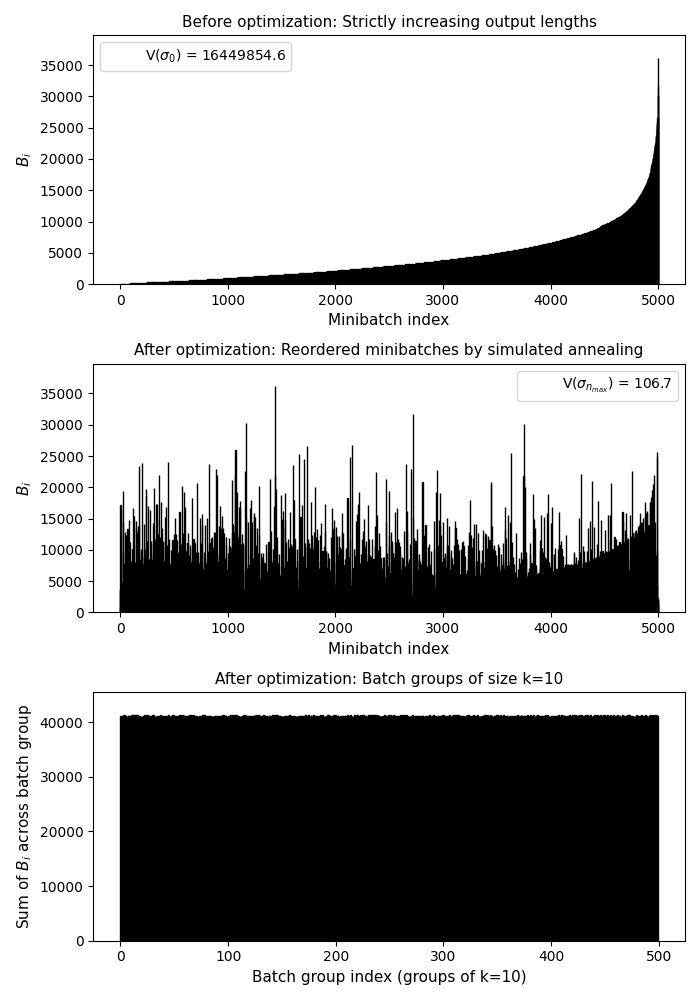}
    \caption{Balanced batch-group ordering. Predicted minibatch loads before (top) and after (middle) simulated annealing, and the resulting group loads for $k=10$ (bottom). Reordering preserves minibatch composition while balancing the load across consecutive groups.}
    \label{fig:app_simannealing}
\end{figure}

Let $B_i$ be the predicted load of minibatch $i$, defined as the sum of its
predicted output lengths. Let $k$ be the number of minibatches per batch group,
and let $G=\lceil N_B/k\rceil$ be the number of groups. The target group load is
\[
M
=
\frac{1}{G}
\sum_{i=1}^{N_B} B_i .
\]
ASP seeks a permutation $\sigma$ minimizing the variance of predicted load
across consecutive groups:
\[
\min_{\sigma \in \mathfrak{S}_{N_B}}
\frac{1}{G}
\sum_{g=1}^{G}
\left(
\sum_{r=(g-1)k+1}^{\min(gk,N_B)}
B_{\sigma(r)}
-
M
\right)^2 .
\]
This objective smooths the predicted generation rate seen by the Training Queue. Instead of sending all short minibatches first and all long minibatches later, ASP mixes minibatches so that each consecutive group has approximately the same predicted load, as illustrated in Figure~\ref{fig:app_simannealing}.


We optimize this objective using simulated annealing~\cite{kirkpatrick1983optimization}, as detailed in Algorithm~\ref{alg:sa}.
The method starts from an initial permutation and repeatedly proposes swaps
between two minibatches. A swap is accepted if it improves the objective, or
with a temperature-dependent probability if it worsens the objective. This
allows the search to escape poor local minima.

\begin{algorithm}[t]
\caption{Simulated annealing for balanced batch-group ordering}
\label{alg:sa}
\begin{algorithmic}[1]
\State \textbf{Input:} minibatch loads $B_1,\ldots,B_{N_B}$, group size $k$, iterations $n_{\max}$, constant $C$
\State Initialize permutation $\sigma_0$
\State Define objective $V(\sigma)$ as the batch-group load variance
\For{$n=0$ \textbf{to} $n_{\max}-1$}
    \State Sample two distinct indices $i,j$ uniformly at random
    \State Let $\sigma'$ be the permutation obtained by swapping $\sigma_n(i)$ and $\sigma_n(j)$
    \State Set temperature $T_n \gets C/\log(n+2)$
    \State Sample $U \sim \mathrm{Uniform}(0,1)$
    \If{$U < \min\left(1,\exp\left((V(\sigma_n)-V(\sigma'))/T_n\right)\right)$}
        \State $\sigma_{n+1}\gets \sigma'$
    \Else
        \State $\sigma_{n+1}\gets \sigma_n$
    \EndIf
\EndFor
\State \textbf{Output:} optimized permutation $\sigma_{n_{\max}}$
\end{algorithmic}
\end{algorithm}

\subsection{Hyperparameters and runtime}
\label{app:scheduler_hyperparameters}

\paragraph{Hyperparameters.} In our experiments, $\kappa$ is kept small, typically in
$\{1,2,3\}$, to prevent excessively large inference minibatches while still
allowing the dynamic program to avoid poor boundaries. The batch-group size
$k$ controls the time scale over which generation load is smoothed. Larger
values of $k$ reduce group-load variance more aggressively, but can also allow larger within-group queue fluctuations. We therefore use intermediate values, typically between $4$ and $10$. Unless otherwise specified, all scheduled runs use base minibatch size $b=32$, $\kappa=2$, and $k=6$, and Algorithm~\ref{alg:sa} runs for $n_{\max}=2\times10^6$ iterations with temperature constant $C=1$.

\paragraph{Runtime.} All three stages of the scheduling pipeline are performed offline, introducing no online scheduling overhead during training. Moreover, the preprocessing cost is negligible compared to the overall training time, even for large datasets.

For step 1, output-length prediction scales linearly with the number of prompts for a fixed predictor. This step is computationally inexpensive, as the predictor is substantially smaller than the model being trained and only needs to generate a few tokens encoding the predicted output length. For step 2, minibatch construction requires sorting the predicted lengths in $O(n\log n)$ time, followed by dynamic programming with $O(\kappa n)$ time complexity, as established in Algorithm~\ref{alg:dp}, which allows for efficient scaling to large datasets.

Finally, in step 3, simulated annealing is highly scalable: each proposed swap affects at most two batch-group loads, allowing the objective to be updated in constant time, independently of $N_B$. To characterize its runtime, we evaluate SA across four synthetic output-length distributions---lognormal, exponential, Pareto, and bimodal---covering $N_B$ up to $10^7$ and $n_{\max}$ up to $10^9$ (Table~\ref{tab:sa_scaling}). At fixed $N_B=100{,}000$, runtime scales approximately linearly with the iteration budget, achieving around 30 million iterations per second with less than 3\% variation across distributions. On a single CPU core, $2.7\times10^5$ iterations already achieve over 95\% variance reduction across all four distributions in under 0.01\,s on average, while even the largest tested budget of $10^9$ iterations completes in just 35.5\,s. At a fixed budget of $2\times10^6$ iterations, runtime remains below one second across all tested minibatch counts, reaching 0.31\,s at $N_B=10^7$. Further improvements are possible through stratified initialization, such as snake-deal assignment, but seemed unnecessary given the already low observed runtimes.


\begin{table*}[t]
\centering
\small
\setlength{\tabcolsep}{5pt}
\caption{%
  SA scaling results across four output-length distributions
  (lognormal, exponential, Pareto, bimodal; individual lengths clipped at 8{,}192 tokens), $k=10$. Variance reduction: $\bigl(1 - V(\sigma_{n_{\max}})/V(\sigma_0)\bigr)\times 100\%$.
  \textbf{Left}: fixed $N_B = 100{,}000$, varying $n_{\max}$.
  Wall time is averaged across distributions (variation $<3\%$).
  \textbf{Right}: fixed $n_{\max} = 2\times10^6$, varying $N_B$.}
\label{tab:sa_scaling}
%
\begin{tabular}{r r r r r r}
\toprule
& & \multicolumn{4}{c}{Variance reduction (\%)} \\
\cmidrule(lr){3-6}
$n_{\max}$ & Wall (s) & Logn. & Exp. & Pareto & Bim. \\
\midrule
$10^{4}$                 & $<0.01$ & 30.6 &  29.8 & 30.5 &  29.6 \\
$8.9\times 10^{4}$       & $<0.01$ & 93.9 &  95.6 & 92.2 &  95.6 \\
$2.7\times 10^{5}$       & $<0.01$ & 98.9 &  99.9 & 97.2 &  99.8 \\
$8.0\times 10^{5}$       &   $0.03$ & 99.4 & 100.0 & 97.6 & 100.0 \\
$2.4\times 10^{6}$       &   $0.09$ & 99.4 & 100.0 & 97.8 & 100.0 \\
$1.2\times 10^{7}$       &   $0.45$ & 99.5 & 100.0 & 97.8 & 100.0 \\
$1.1\times 10^{8}$       &    $4.0$ & 99.5 & 100.0 & 97.8 & 100.0 \\
$10^{9}$                 &   $35.5$ & 99.5 & 100.0 & 97.8 & 100.0 \\
\bottomrule
\end{tabular}
\hspace{2em}
%
\begin{tabular}{r r}
\toprule
$N_B$ & Wall (s) \\
\midrule
$10^{2}$                 & $0.05$ \\
$896$                    & $0.05$ \\
$4{,}642$                & $0.08$ \\
$24{,}041$               & $0.07$ \\
$1.2\times 10^{5}$       & $0.06$ \\
$6.4\times 10^{5}$       & $0.11$ \\
$1.9\times 10^{6}$       & $0.16$ \\
$5.8\times 10^{6}$       & $0.28$ \\
$10^{7}$                 & $0.31$ \\
\bottomrule
\end{tabular}
\end{table*}

\section{Experimental details}
\label{app:experimental_details}

This appendix provides additional details for the experiments in
Section~\ref{sec:results}: hardware, datasets, length prediction, node
allocations, and additional plots.

\subsection{Hardware and implementation}

\paragraph{Hardware.}
All training and generation experiments are conducted on nodes that are each equipped with 8 Ascend 910B3 NPUs. The simulated annealing runtime benchmarks in Appendix~\ref{app:scheduling_details} are conducted on a single CPU core.
\paragraph{ASP implementation.}
ASP is implemented on top of MindSpeed-RL using GRPO, with vLLM (v0.9.1) as the generation backend. The weight-flow mechanism described in Appendix~\ref{app:weightflow} handles the conversion of trainer snapshots into inference-compatible weights. The scheduling steps are implemented as per the pseudocodes given in Appendix~\ref{app:scheduling_details}.

\paragraph{Baselines.}
We compare ASP against two baselines. The first is the original MindSpeed-RL hybrid pipeline, which serves as backend framework for ASP. This same-hardware comparison captures the combined effect of ASP's architectural and implementation changes, while the specific contribution of scheduling is isolated in Section~\ref{sec:scheduling_ablation}.

The second baseline is verl (v0.7.1), with vLLM for generation and FSDP~\cite{zhao2023pytorchfsdp} for training. We use its one-step off-policy pipeline, which generates the next rollout batch while updating the policy on the current batch.

\subsection{Models and datasets}

Our main end-to-end experiments evaluate two Qwen3 models: Qwen3-30B-A3B, a Mixture-of-Experts model, and Qwen3-32B, a dense model~\cite{qwen3-techreport}. We additionally use Qwen3-8B for scheduling ablations and predictor-sensitivity experiments.

We use two datasets. \textsc{Deepscaler}~\cite{deepscaler2025} contains
approximately 41K mathematical problems and is used for throughput,
scheduling, and scalability experiments. \textsc{Maze-Reasoning-GRPO-v0.1}
contains more than 180K maze problems and is used for convergence evaluation
\cite{dao2025alphamazeenhancinglargelanguage}. We use the maze task because it
is less likely to be saturated by the pre-trained model than standard math problems
already represented in modern LLM training data.

For the maze benchmark, we report two rewards:
\emph{maze integrity}, which measures whether the generated solution is
well-formed, and \emph{maze correctness}, which measures whether the decoded
path solves the maze. These rewards are used only to compare learning dynamics
under ASP and the hybrid baseline.

\subsection{Output-length distributions}

\begin{figure}[t]
    \centering
    \begin{subfigure}[t]{0.48\linewidth}
        \centering
        \includegraphics[width=\linewidth]{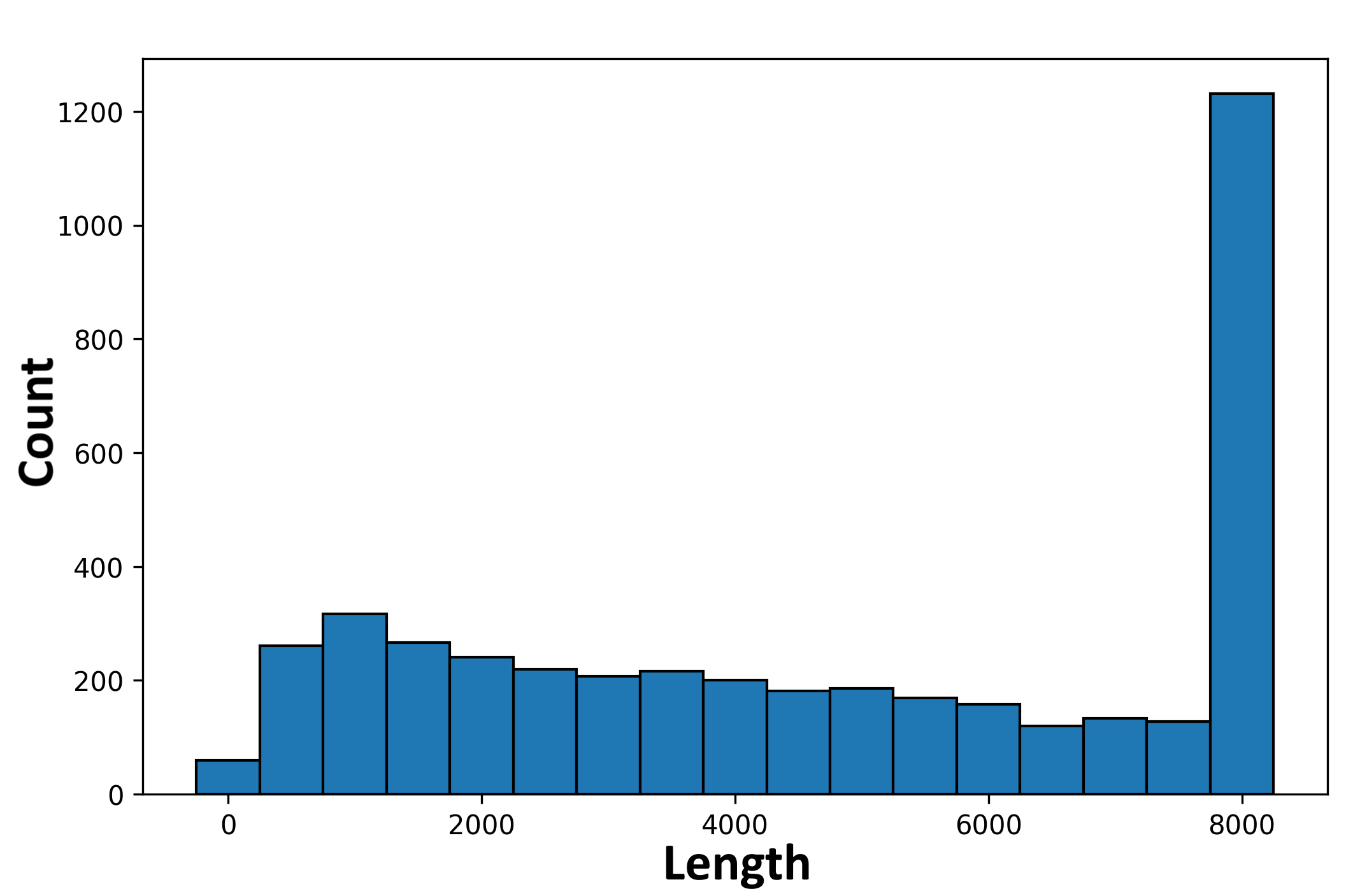}
        \caption{\textsc{Deepscaler}, capped at 8192 tokens.}
        \label{fig:app_output_distribution_deepscaler}
    \end{subfigure}
    \hfill
    \begin{subfigure}[t]{0.48\linewidth}
        \centering
        \includegraphics[width=\linewidth]{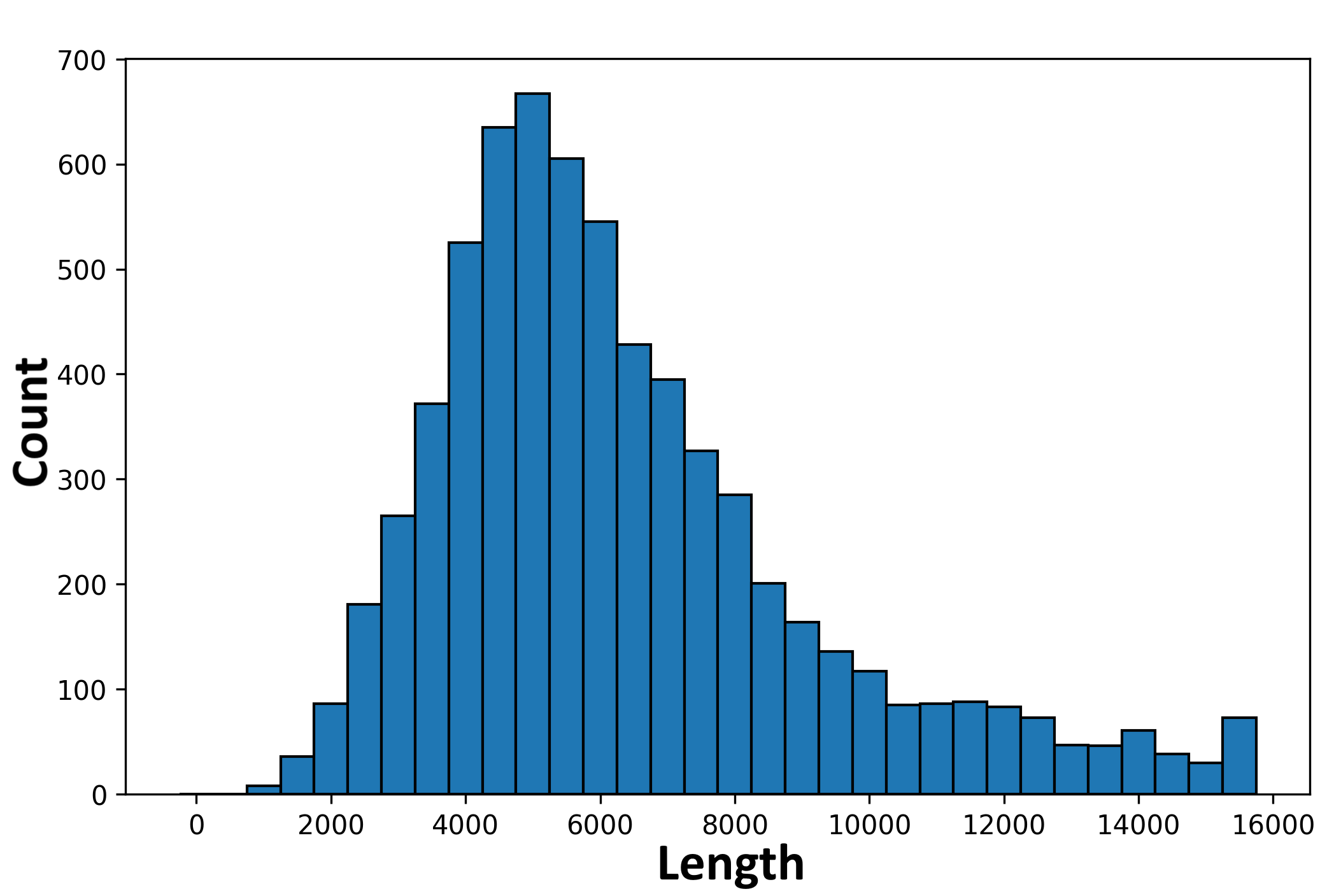}
        \caption{\textsc{Maze-Reasoning-GRPO}, capped at 16384 tokens.}
        \label{fig:app_output_distribution_maze}
    \end{subfigure}
    \caption{Output-length distributions for the datasets used in our
    experiments. The long tails motivate length-aware scheduling.}
    \label{fig:app_output_distribution}
\end{figure}

The scheduling problem is motivated by long and variable output lengths.
Figure~\ref{fig:app_output_distribution} shows the output-length distributions
used in our experiments. For \textsc{Deepscaler}, we set the maximum output
length to 8192 tokens. A significant fraction of outputs reaches this cap,
suggesting that longer contexts may further increase the relative importance of
rollout scaling. For \textsc{Maze-Reasoning-GRPO-v0.1}, we use a maximum output
length of 16384 tokens.

\subsection{Output-length predictor}
\label{app:output_length_predictor}

The output-length predictor is fine-tuned on 4000 prompt--length pairs extracted
from the target datasets. The cost of fine-tuning this predictor is negligible
relative to the target RL training run. The predictor remains fixed throughout
training and is used only during the offline scheduling pass.

We evaluate the predictor using Kendall's rank correlation coefficient
$\tau$~\cite{kendall1938rank}, because ASP's sorting mechanism in the first step relies primarily on relative order rather than exact length values. For \texttt{Qwen3-30B-A3B} on \textsc{Deepscaler}, the predictor reaches
approximately $\tau=0.64$ with an 8192-token cap and $\tau=0.55$ with a
16384-token cap. The value $\tau=0.64$ corresponds to roughly 82\% of prompt
pairs being ordered consistently with the measured output lengths, which is a
substantial improvement over random ordering.

\subsection{Configurations for performance and scalability}
\label{app:run_configurations}


\subsubsection{Matched-resource comparisons}



The main comparisons in Table~\ref{tab:asp-vs-baseline} use eight nodes (64 NPUs) per system. ASP and verl each allocate 32 NPUs to training and 32 to generation, whereas the MindSpeed-RL hybrid baseline alternates training and rollout on the same 64-NPU pool. 
For verl, generation uses tensor parallelism of 8 and vLLM continuous batching. Training uses remove-padding (\texttt{use\_remove\_padding=True}), dynamic batch sizing for actor/reference passes (\texttt{use\_dynamic\_bsz=True}), gradient checkpointing, and CPU offloading of parameters and optimizer states.

For \texttt{Qwen3-30B-A3B}, each iteration processes 16 prompts with eight responses per prompt, giving 128 trajectories. For \texttt{Qwen3-32B}, it processes 32 prompts with eight responses per prompt, giving 256 trajectories. For each model, verl uses the same global batch size and prompt and response token limits as the corresponding ASP run. 

Table~\ref{tab:app_run_configurations} details the ASP and MindSpeed-RL layouts and measurement windows. 

\begin{table}[t]
  \centering
  \caption{ASP and MindSpeed-RL configurations for Table~\ref{tab:asp-vs-baseline}. TP, PP, DP, and EP denote tensor, pipeline, data, and expert parallelism, respectively. All listed trainer layouts use context parallelism degree 1. Generation parallelism is specified per ASP actor or for the MindSpeed-RL rollout layout. A dash indicates no dedicated generation pool; N/A indicates that expert parallelism does not apply.}
  \label{tab:app_run_configurations}
  \small
  \setlength{\tabcolsep}{4pt}
  \begin{tabular}{@{}lcccc@{}}
    \toprule
      & \multicolumn{2}{c}{\texttt{Qwen3-30B-A3B}}
      & \multicolumn{2}{c}{\texttt{Qwen3-32B}} \\
    \cmidrule(lr){2-3}\cmidrule(lr){4-5}
    Setting & ASP & MindSpeed-RL & ASP & MindSpeed-RL \\
    \midrule
    \multicolumn{5}{@{}l}{\textit{Workload per iteration}} \\
    Prompts & 16 & 16 & 32 & 32 \\
    Responses per prompt & 8 & 8 & 8 & 8 \\
    Trajectories & 128 & 128 & 256 & 256 \\
    \addlinespace
    \multicolumn{5}{@{}l}{\textit{Training and generation}} \\
    Training NPUs & 32 & 64 & 32 & 64 \\
    Dedicated generation NPUs & 32 & --- & 32 & --- \\
    Trainer TP / PP / DP & 4 / 1 / 8 & 4 / 1 / 16 & 8 / 2 / 2 & 8 / 4 / 2 \\
    Trainer EP & 4 & 8 & 1 & 1 \\
    Dedicated generation actors & 16 & --- & 16 & --- \\
    Generation TP / PP & 2 / 1 & 4 / 1 & 2 / 1 & 4 / 1 \\
    Generation EP & 1 & 8 & N/A & N/A \\
    \addlinespace
    \multicolumn{5}{@{}l}{\textit{Measurement (including startup)}} \\
    Iterations & 120 & 20 & 50 & 50 \\
    Mean iteration time (s) & 215.2 & 927 & 230.7 & 928 \\
    \bottomrule
  \end{tabular}
\end{table}

\subsubsection{Scalability}

The \texttt{Qwen3-30B-A3B} scalability series keeps the workload fixed at 16 prompts and eight responses per prompt (128 trajectories per iteration) for ASP, MindSpeed-RL, and verl at every tested scale. The verl baseline uses the same settings as in the matched-resource comparison, except that dynamic batch sizing is disabled. Table~\ref{tab:app_node_allocations} maps total resources to ASP's training/generation split and to the generation-NPU counts used in Figure~\ref{fig:app_scalability_comparison}.


\begin{table}[t]
  \centering
  \caption{Resource allocations for the \texttt{Qwen3-30B-A3B} scalability series.}
  \label{tab:app_node_allocations}
  \small
  \setlength{\tabcolsep}{10pt}
  \begin{tabular}{@{}rrr@{}}
    \toprule
    Total NPUs & ASP training NPUs & ASP generation NPUs \\
    \midrule
    16  & 8  & 8 \\
    32  & 8  & 24 \\
    64  & 32 & 32 \\
    128 & 48 & 80 \\
    \bottomrule
  \end{tabular}
\end{table}




\subsection{Scheduling ablation details}
\label{app:scheduling_ablation}

This appendix provides additional details on the scheduling experiments presented in Section~\ref{sec:scheduling_ablation}: the full-training comparison with and without scheduling, the robustness analysis across prediction accuracies, and the evaluation with real generation and simulated training. We report the experimental configurations, methodology, and additional results for each.

\subsubsection{Training with and without scheduling}
\label{app:scheduling_onoff}

\begin{figure}[t]
  \centering
  \includegraphics[width=\linewidth]{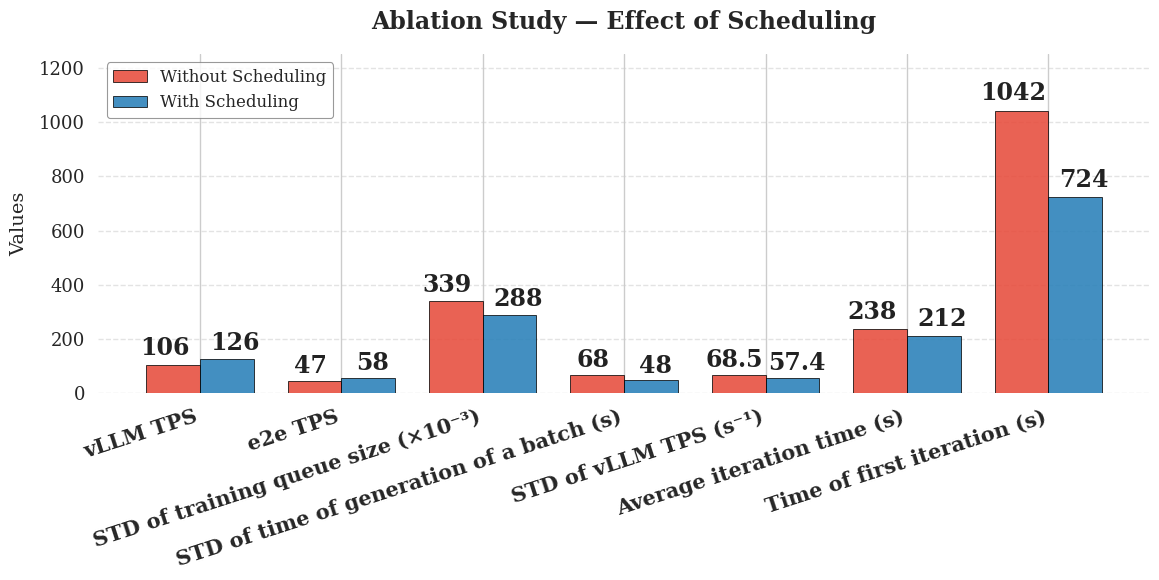}
  \caption{Training with and without scheduling on \texttt{Qwen3-30B-A3B} using \textsc{Deepscaler}, with 200 iterations per run. vLLM TPS measures generated non-padding tokens per second per generation NPU; end-to-end (e2e) TPS uses full iteration time and all NPUs. Mean iteration time includes startup, which is also reported separately. Queue STD is reported in $10^{-3}$ global batches.}
  \label{fig:schedablation}
\end{figure}

Figure~\ref{fig:schedablation} compares training with and without scheduling on \texttt{Qwen3-30B-A3B} using \textsc{Deepscaler}. Both runs use the allocation, trainer/generation parallelism, and 16-prompt workload of the main MoE ASP run in Table~\ref{tab:app_run_configurations}, with eight responses per prompt. Prompt and response limits are 1024 and 8192 tokens, respectively. Both runs last 200 training iterations and differ only in whether scheduling is enabled. The scheduled run uses the default scheduler settings (Appendix~\ref{app:scheduler_hyperparameters}).

Enabling scheduling reduces mean iteration time from 238 to 212\,s (10.9\%) and increases end-to-end throughput from 47 to 58 tokens/s/NPU (23.4\%). 
Scheduling also reduces the standard deviation (STD) of
minibatch generation time from 68 to 48\,s (29.4\%) and
that of Training Queue occupancy by 15.0\%, indicating more consistent generation times and queue occupancy.

Considering generation alone, vLLM throughput, measured in generated non-padding tokens per second per generation NPU, increases from 106 to 126 tokens/s/NPU, while its STD decreases from 68.5 to 57.4 tokens/s/NPU. Unlike end-to-end throughput, this metric focuses on generation rather than the full training pipeline. Mean iteration time includes all 200 iterations, including
the first (startup) iteration, which is also reported
separately and decreases from 1042 to 724\,s (30.5\%).

\subsubsection{Robustness of the scheduler to output length prediction quality}
\label{app:scheduling_sweeps}

Here, we study the robustness of the scheduling pipeline to the quality of the output length predictions. 

\paragraph{Conditions and fixed responses.}
We use 39{,}000 measured responses from each of \texttt{Qwen3-32B} and
\texttt{Qwen3-8B}. For each model, a complete inference pass supplies fixed
measured lengths $\ell_i$. C0 forms fixed-size minibatches in original
order; C1 sorts by predicted length before fixed-size batching; C2 applies
dynamic programming to refine the sorted minibatch boundaries;
C3 randomly shuffles the C2 minibatches; and C4 applies ASP's balanced ordering.
C2--C4 have identical minibatch composition. Their local length metrics
are therefore identical, while their batch-group loads can differ.

\paragraph{Scheduler configuration.}
Both models use the default scheduler settings (Appendix~\ref{app:scheduler_hyperparameters}).
C0 and C1 use fixed-size minibatches of 32; C2--C4 allow sizes from 32 to 34. 
The experiment with simulated training in Appendix~\ref{app:scheduling_replay} uses the same scheduler settings.

\paragraph{Prediction perturbations.}
Holding real output lengths $\ell_i$ fixed, we construct simulated predictions using
\[
  \widehat{\ell}_i=\ell_i\exp(\nu\varepsilon_i),
  \qquad \varepsilon_i\sim\mathcal{N}(0,1),
\]
calibrating $\nu$ to achieve a target Kendall correlation 
$\tau(\widehat{\ell},\ell)$. For each quality level and seed, all conditions
receive the same prediction vector. We evaluate schedule metrics on the
measured lengths, not on the predictions, and report means over three seeds.
The nominal quality levels range from the oracle condition ($\tau=1$)
to the random-quality endpoint ($\tau=0$). Because generation is stochastic,
the oracle uses information from the measured responses and is an upper
bound rather than an attainable predictor operating point.

The measured ranker correlations are 0.64 for \texttt{Qwen3-30B-A3B},
0.44 for \texttt{Qwen3-32B}, and 0.58 for \texttt{Qwen3-8B}.

\paragraph{Metrics.}
For the $j$th minibatch, let $\mathcal{I}_j$ denote its
set of prompt indices, and let $\bar{\ell}_j$ and
$\ell_{\max,j}$ denote its mean and maximum measured
response lengths. With $N_B$ minibatches, we report

\[
  D=\frac{1}{N_B}\sum_{j=1}^{N_B}
       \operatorname{STD}\{\ell_i:i\in \mathcal{I}_j\},
  \qquad
  P=\frac{1}{N_B}\sum_{j=1}^{N_B}
       \left(1-\frac{\overline{\ell}_j}{\ell_{\max,j}}\right).
\]

Here, $D$ measures the average length dispersion within minibatches (in tokens), while $P$ denotes the mean padding rate.

To assess load balance across consecutive batch groups, let $L_g$ denote the total number of measured response tokens in group $g$. We report the coefficient of variation (CV), defined as $\operatorname{STD}(L_g)/\operatorname{mean}(L_g)$, and the group-load spread, defined as $\max_g L_g/\min_g L_g$.

Finally, we report prediction mean absolute error (MAE) in tokens alongside Kendall's $\tau$.

\begin{table}[t]
  \centering
  \caption{Local schedule metrics across controlled predictor qualities on three seeds. C2--C4 share
  minibatches and therefore share dispersion and padding rate. }
  \label{tab:app_scheduling_local}
  \small
  \setlength{\tabcolsep}{4pt}
  \begin{tabular}{@{}rrrrrrrr@{}}
    \toprule
    & & \multicolumn{3}{c}{Dispersion (tokens)}
      & \multicolumn{3}{c}{Mean padding rate} \\
    \cmidrule(lr){3-5}\cmidrule(lr){6-8}
    $\tau$ & MAE & C0 & C1 & C2--C4 & C0 & C1 & C2--C4 \\
    \midrule
    \multicolumn{8}{@{}l}{\texttt{Qwen3-32B}} \\
    1.00 &    0 & 2138 &    5 &    5 & 0.322 & 0.003 & 0.003 \\
    0.75 &  839 & 2138 &  820 &  828 & 0.322 & 0.176 & 0.175 \\
    0.64 & 1203 & 2138 & 1163 & 1163 & 0.322 & 0.226 & 0.224 \\
    0.50 & 1692 & 2138 & 1558 & 1563 & 0.322 & 0.273 & 0.271 \\
    0.25 & 2729 & 2138 & 2064 & 2057 & 0.322 & 0.330 & 0.328 \\
    0.00 & 3826 & 2138 & 2267 & 2253 & 0.322 & 0.348 & 0.347 \\
    \addlinespace
    \multicolumn{8}{@{}l}{\texttt{Qwen3-8B}} \\
    1.00 &    0 & 2066 &    5 &    5 & 0.294 & 0.003 & 0.003 \\
    0.75 &  819 & 2066 &  783 &  783 & 0.294 & 0.154 & 0.152 \\
    0.64 & 1186 & 2066 & 1123 & 1123 & 0.294 & 0.206 & 0.204 \\
    0.50 & 1682 & 2066 & 1525 & 1525 & 0.294 & 0.250 & 0.251 \\
    0.25 & 2712 & 2066 & 1998 & 1998 & 0.294 & 0.296 & 0.295 \\
    0.00 & 3879 & 2066 & 2190 & 2190 & 0.294 & 0.313 & 0.313 \\
    \bottomrule
  \end{tabular}
\end{table}

\begin{table}[t]
  \centering
  \caption{Batch-group load CV for the same controlled sweeps as
  Table~\ref{tab:app_scheduling_local}. Entries are three-seed means,
  evaluated on measured response lengths.}
  \label{tab:app_scheduling_balance}
  \small
  \setlength{\tabcolsep}{9pt}
  \begin{tabular}{@{}rrrrrr@{}}
    \toprule
    $\tau$ & C0 & C1 & C2 & C3 & C4 \\
    \midrule
    \multicolumn{6}{@{}l}{\texttt{Qwen3-32B}} \\
    1.00 & 0.103 & 0.495 & 0.492 & 0.188 & 0.000 \\
    0.75 & 0.103 & 0.447 & 0.445 & 0.175 & 0.052 \\
    0.64 & 0.103 & 0.403 & 0.401 & 0.162 & 0.070 \\
    0.50 & 0.103 & 0.332 & 0.330 & 0.140 & 0.080 \\
    0.25 & 0.103 & 0.183 & 0.183 & 0.115 & 0.092 \\
    0.00 & 0.103 & 0.097 & 0.111 & 0.088 & 0.091 \\
    \addlinespace
    \multicolumn{6}{@{}l}{\texttt{Qwen3-8B}} \\
    1.00 & 0.097 & 0.458 & 0.455 & 0.172 & 0.000 \\
    0.75 & 0.097 & 0.416 & 0.414 & 0.159 & 0.049 \\
    0.64 & 0.097 & 0.375 & 0.374 & 0.149 & 0.062 \\
    0.50 & 0.097 & 0.309 & 0.308 & 0.131 & 0.079 \\
    0.25 & 0.097 & 0.171 & 0.172 & 0.105 & 0.088 \\
    0.00 & 0.097 & 0.091 & 0.094 & 0.085 & 0.083 \\
    \bottomrule
  \end{tabular}
\end{table}

\paragraph{Interpretation.}
Table~\ref{tab:app_scheduling_local} shows that sorting's local benefit
decreases as prediction quality deteriorates and is already marginally negative at $\tau=0.25$ on both models. The dynamic programming step of ASP changes mean padding by at most 0.002
relative to C1 across these sweeps, and does not improve every condition.
Its small average effect reflects changes to only a few boundaries among
approximately 1{,}200 minibatches. Individual boundaries can nevertheless
be costly: for models such as \texttt{Qwen3-30B-A3B}, we observed that misplaced boundaries could delay the generation of one minibatch by up to 12 minutes.

Table~\ref{tab:app_scheduling_balance} shows the complementary ordering
effect. For $\tau\geq0.50$, C1/C2 have approximately three to five times
C0's load CV. C4 keeps CV at or below C0 throughout the tested sweep and
outperforms random shuffling at every tested level with $\tau\geq0.25$.
For $\tau=0$, on the 32B model, C3 is slightly lower than C4
(0.088 versus 0.091); both remain below C0. Thus, the results support
robust load balance in this sweep.


\subsubsection{Real generation with simulated training}
\label{app:scheduling_replay}

As described in Section~\ref{sec:scheduling_ablation}, we evaluate C0--C4 using real \texttt{Qwen3-8B} generation with modeled training times. All five configurations use the same prompts, predictions, resources, token limits, and architecture. The predictor has a measured Kendall correlation of $\tau=0.58$.

By keeping training times fixed, we can attribute differences in generation times, sample arrivals, and Training Queue behavior to scheduling alone.

To justify using fixed training times, we measure
the variability of training-step durations in an ASP
run with one TU node and three GU nodes. Over 19
post-warmup iterations, training-step durations have
a coefficient of variation (CV) of only 5.0\%,
compared with 6.8\% for end-to-end iteration times.
This limited variability supports using fixed
training durations to isolate the effects of
rollout scheduling.

C4 achieves the lowest mean iteration time, 178\,s versus 188\,s for C0
(5.3\% lower). The iteration-index/time Spearman correlations are
$-0.080$, $+0.670$, $+0.510$, $+0.190$, and $-0.075$ for C0--C4,
respectively. C4 therefore shows little to no monotonic drift, whereas the
sorted C1/C2 schedules become considerably slower over successive iterations. Shuffling matches
C4's queue STD (36 samples) and has a slightly lower queue maximum
(104 versus 110 samples), but a higher mean time (190\,s).



\subsection{Main-run timing, queue, and response-length diagnostics}
\label{app:scheduling_runtime_traces}

\begin{figure}[t]
  \centering
  \includegraphics[width=\linewidth]{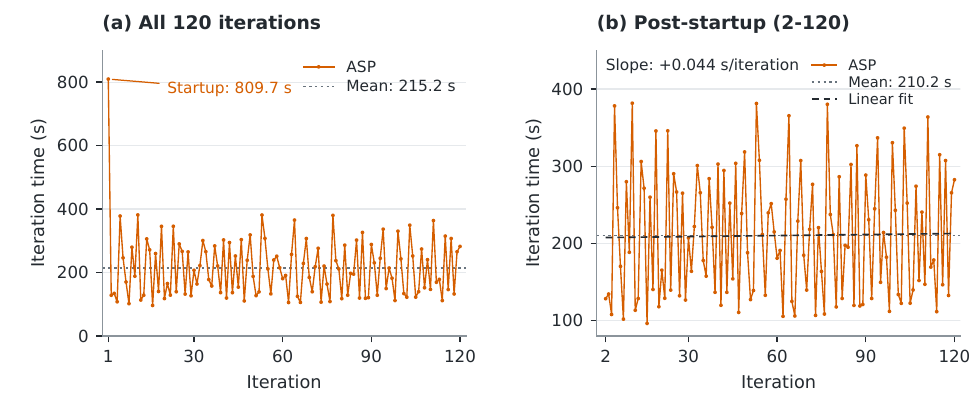}
  \caption{Iteration times for the complete \texttt{Qwen3-30B-A3B} run underlying Table~\ref{tab:asp-vs-baseline}. (a) All 120 iterations, including the 809.7\,s startup; the mean is 215.2\,s. (b) Iterations 2--120, averaging 210.2\,s; the dashed linear fit has slope $+0.044$\,s per iteration.}
  \label{fig:app_iteration_trace}
\end{figure}

We additionally report runtime diagnostics from the complete
\texttt{Qwen3-30B-A3B} experiment underlying
Table~\ref{tab:asp-vs-baseline}. Its 215.2\,s mean includes all 120
iterations, including the 809.7\,s first iteration. The remaining 119
iterations average 210.2\,s, with a fitted linear slope of
$+0.044$\,s per iteration and 20-iteration means between 197.0 and
221.2\,s. These summaries indicate little sustained timing drift over
the observed run.
Figure~\ref{fig:app_iteration_trace} shows the complete trace and its post-startup variation.

Across 16 generation actors, recorded prompt counts range from 112
to 140, with a CV of 6.4\%. Queue depth measured at arrivals averages
0.39 global batches, with a 95th percentile of 0.75 and a maximum of
1.25 global batches, without progressive accumulation. 



Across the 1{,}992 recorded prompts, mean response length increases from 5{,}637 tokens at policy versions 1--20 to 5{,}730 at versions 100--115 (1.6\%). A linear trend in mean response
length across 103 recorded versions has $R^2=0.013$.
This supports limited mean-length drift in the observed run. Longer runs, curricula, or domain shifts may require renewed profiling or predictor updates.

\subsection{Reproducibility notes}

The experiments require access to Ascend 910B3 nodes, MindSpeed-RL, the Qwen3 model checkpoints, the datasets listed above, and the ASP scheduling and
weight-flow implementation. The main sources of variability are the sampled prompts, generation stochasticity, scheduling predictor quality, and system load during generation and weight refresh. For reproducibility, we record the model, dataset, output-length cap, node allocation, scheduler configuration
$(b,\kappa,k)$, measured Kendall's $\tau$, generation parameters, and measured timing breakdown for each run.

Appendix~\ref{app:run_configurations} reports the performance and scalability configurations; Appendix~\ref{app:scheduling_onoff}--\ref{app:scheduling_replay} give the ablation settings and protocols. Supporting ASP timing and queue diagnostics are reported in Appendix~\ref{app:scheduling_runtime_traces}.

\section{Limitations and discussion}
\label{app:limitations}

ASP has several limitations, which suggest directions for future improvement.
First, our prototype is implemented in the
MindSpeed-RL ecosystem on Ascend 910B3 NPUs. 
This gives a controlled same-hardware comparison against the hybrid pipeline from which ASP is derived, and our ablations isolate the effect of the scheduling mechanism. Direct comparisons with StreamRL~\citep{streamRL},
AsyncFlow~\citep{asyncflow},
AReaL~\citep{fu2026areallargescaleasynchronousreinforcement},
and Laminar~\citep{sheng2025laminarscalableasynchronousrl}
were not conducted, as establishing comparable
configurations on our specific Ascend NPU stack would require
substantive integration and validation. Such comparisons
could confound system differences with implementation quality. We partially address this limitation by adding a verl-based disaggregated baseline in Section~\ref{sec:results}, as verl is an actively maintained open-source RL post-training framework that supports separated rollout and training configurations~\cite{verl,hybridflow}.
This comparison should be interpreted as measuring the benefit of ASP over a disaggregated implementation without ASP's scheduling layer.

Second, the output-length predictor is fixed during training. This design keeps ASP's scheduling pass simple and avoids online overhead, while still delivering scheduling gains in our runs. 
Longer training runs, stronger
policy shifts, or curriculum changes may benefit from periodically refreshing the predictor or combining offline scheduling with lightweight online correction.

Third, our main end-to-end evaluations focus on Qwen3-30B-A3B and Qwen3-32B, complemented by scheduling ablations and predictor-sensitivity experiments on Qwen3-8B. Larger models, longer context windows, and workloads with more extreme output-length distributions may amplify
the benefits of multi-instance rollout and scheduling, but they
should be validated directly. Conversely, short-output workloads may see smaller
gains because rollout variability is less dominant.

Fourth, our available NPU resources are limited to 128 Ascend NPUs in our experiments.
Thus, our scalability test stops at the 128-NPU point.
Although the trend is already clear, larger-scale experiments would provide a stronger evaluation, especially for studying the impact of crossing super-node regions or leaf switches.

Finally, ASP assumes that the prompt set is known before rollout begins, since
the scheduler performs an offline preprocessing pass. This matches many RL
post-training settings where prompts are sampled from a fixed dataset, but it is
less directly applicable to fully online data-generation pipelines. Extending
the same ideas to streaming prompts would require an online version of
length-aware minibatching and balanced dispatch.

Beyond the controlled experiments reported above, ASP has also been used in an
internal telecom-domain RL post-training workflow for \texttt{Qwen3-30B-A3B}. In
that deployment, the overall training period was reduced from more than three
months to about two months, and the resulting domain-enhanced model was used for
telecom applications such as knowledge answering and tool calling. Because the
workflow, data, and deployment details are confidential, we report this only as
qualitative evidence of practical applicability, not as part of the controlled
experimental comparison.

\section{Generative AI assistance}
\label{sec:AI_usage}
We used generative AI tools to assist with writing, translation, revision, and language polishing; identifying and discussing relevant literature; and code implementation and debugging. These tools were not used for research ideation, methodology design, or conducting or analyzing the experiments. The authors reviewed the AI-assisted material and take full responsibility for the final manuscript, code, and reported results.


\end{document}